\documentclass[final]{siamltex}

\usepackage{amssymb,amsmath,graphicx}

\def\TAN {{\tilde{A}}_N}

\newcommand{\RR}{{\mathbb{R}}}
\newcommand{\CC}{{\mathbb{C}}}

\newtheorem{remark}{Remark}

\title{A Krylov Stability-Corrected Coordinate-Stretching Method to Simulate Wave Propagation in Unbounded Domains}
\author{Vladimir Druskin\thanks{Schlumberger Doll Research, 1 Hampshire St., Cambridge, MA 02139, US ({\tt Druskin1@slb.com}).}
        \and Rob Remis\thanks{Circuits and Systems Group, Faculty of Electrical Engineering, Mathematics and Computer Science, 
Delft University of Technology, Mekelweg 4, 2628~CD Delft, The Netherlands ({\tt R.F.Remis@tudelft.nl})}}

\begin{document}
\maketitle

%
%
\begin{abstract}
The Krylov subspace projection approach is a well-established tool for the reduced order modeling of dynamical systems in the time domain. 
In this paper, we  address the main issues obstructing the application of this powerful approach to the time-domain solution of exterior wave problems.
We use frequency independent perfectly matched layers to simulate the extension to infinity. Pure imaginary stretching functions based on Zolotarev's optimal rational approximation of the square root are implemented leading to perfectly matched layers with a controlled accuracy over a complete spectral interval of interest. A new Krylov-based solution method via stability-corrected operator exponents is presented which allows us to construct reduced-order models (ROMs) that respect the delicate spectral properties of the original scattering problem. The ROMs are unconditionally stable and are based
on a renormalized bi-Lanczos algorithm. We give a theoretical foundation of our method and illustrate its performance 
through a number of numerical examples in which we simulate 2D electromagnetic
wave propagation in unbounded domains, including a photonic waveguide example. The new algorithm outperforms the conventional finite-difference time domain method for problems on large time intervals.
\end{abstract}

%
%


\pagestyle{myheadings}
\thispagestyle{plain}
\markboth{VLADIMIR DRUSKIN AND ROB REMIS}{KRYLOV STABILITY CORRECTED COORDINATE STRETCHING}

%
%
\section{Introduction}
\label{sec:intro}
The Krylov subspace projection approach is a well-established tool for model reduction of large scale linear dynamical systems \cite{Antoulas}. It is especially efficient when the late time solution can be accurately approximated via a relatively small numbers of eigenmodes as is the case in damped oscillatory problems, for example. In addition, it is well known that under some regularity assumptions, solutions of initial-value problems for homogeneous wave equations in unbounded domains can also be obtained by solving damped problems with energy decaying in any bounded subdomain (even in the case of lossless media). Furthermore, for the case of odd spatial dimensions, the late time evolution of such solutions can be asymptotically expanded via a  sum of time-exponential modes. These modes correspond to so-called scattering resonances or poles, which can be viewed as a surrogate of discrete eigenvalues for exterior problems~\cite{LaxPhillips,TangZworski}. The above observations give us good motivation to extend the Krylov based reduced-order model (ROM) approach to the solution of transient exterior wave problems. 

The main difficulty with applying a model reduction technique to exterior wave problems is that such techniques lead to {\it nonlinear} eigenproblems for spatial dimensions larger than one \cite{LaxPhillips,Taylor}.  This eigenproblem, however, becomes {\it linear} complex symmetric in the framework of the complex scaling method of Aguilar-Balslev-Combes-Simon theory \cite{Hislop&Sigal} introduced in the beginning of the 70s by Aguilar and Combes \cite{Aguilar&Combes}, Balslev and Combes \cite{Balslev&Combes}, and Simon \cite{Simon}. This scaling method is used in atomic and molecular physics to calculate energies, widths, and cross-sections of open quantum systems \cite{Moiseyev}. 

The complex scaling method is equivalent to the Berenger's perfectly matching layer (PML) \cite{Berenger} with frequency independent imaginary damping. 
In particular,  Bindel et al. \cite{Bindel} considered a PML with frequency independent imaginary damping that was optimized for a certain frequency $\omega_0$ and then used projection onto the rational Krylov subspace generated by the Taylor series expansion of the resolvent around that frequency. Such an approach yielded reasonably good frequency domain ROMs in some vicinity of $\omega_0$ (for frequencies different from $\omega_0$ by a factor of two or three). This allowed for the efficient computation of scattering poles in a neighborhood of a point determined by $\omega_{0}$. 
Resonances in open systems have also been computed  using some equivalent PML and  complex scaling method formulations in \cite{Hein_etal1, Hein_etal2, Olyslager} and a theoretical analysis of numerically computing resonances via  the finite element method is given in \cite{Kim&Pasciak}. 

Fixing the PML frequency to $\omega_{0}$ has two main obstacles to become global, however, in particular for time domain field approximations.  First,
the classical PML \cite{Berenger} has frequency-dependent imaginary damping (stretching) that makes attenuation frequency independent. In contrast, attenuation of the fixed frequency PML~(FFPML) is frequency-dependent and its accuracy therefore deteriorates away from $\omega_0$. 
Second, and most importantly, the exact frequency domain solution is a multivalued function with a branch cut on the negative semi-axis and {\it no poles} in the complex plane \cite{LaxPhillips,Taylor}, while the ROM designed in \cite{Bindel} is single valued and meromorphic on the complex plane. Transformed to the time domain, such a ROM produces complex and exponentially unstable solutions \cite{Bindel1}.  

In this paper, we design stable and accurate time domain ROMs by addressing the above mentioned problems. We adopt the optimal PML approximation of \cite{Asvadurov_etal}  based on the Zolotarev  optimal rational approximation of the square root. It yields an FFPML with a stencil of a standard second-order finite difference scheme, but with uniform exponential convergence on a prescribed frequency interval and with a convergence rate only logarithmically dependent on the ratio  of maximum and minimum frequency. Subsequently, we represent wave evolution via stability-corrected time-domain exponential (SCTDE) operator functions. These functions yield unconditionally stable solutions  respecting the delicate spectral properties of the original scattering problem.  We apply a Lanczos-based algorithm to compute the stability corrected exponents of the discretized operators, with a cost per step that is comparable with the cost of one iteration of explicit FDTD using a conventional PML. The new algorithm is targeted to computing wave field evolution for long time intervals, where it significantly outperforms FDTD.

This paper is organized as follows. In Section~\ref{sec:stabcor}, we formulate a model 2D wave problem and introduce the necessary background information on the PML approach. In Section~\ref{sec:discretization} we discretize our exterior wave problem by using a uniform five-point finite-difference scheme with real step sizes in the domain of interest and the exponentially convergent Zolotarev-based FFPML discretization with imaginary step sizes in the exterior. An optimal asymptotic error bound for the Zolotarev-based discretization scheme is given as well. In Section~\ref{SCTDE}, we  present an expression for the solution based on the SCTDE matrix function and derive a Plancherel identity connecting the errors of the SCTDE and the uncorrected frequency domain solutions. A renormalized complex Lanczos algorithm for the computation of the SCTDE matrix function is described in Section~\ref{sec:krylov}. Finally, numerical experiments for electromagnetic wave propagation in 2D unbounded problems are given in Section~\ref{sec:em_2D}. A photonic waveguide example is presented that shows a significant speedup compared with FDTD for large wave propagation times.

\subsection{Notation and principal value convention for the square root}
\label{subsec:convention}

A time-domain field quantity is denoted by $\mathsf{u}$, while its Laplace/Fourier transformed counterpart is written as $u$. A superscript~$N$ is used for the discretized counterparts of continuous field quantities and the subscript $m$ indicates that a field approximation is drawn from an $m$-dimensional Krylov subspace.
For example, $u^{N}$ is the discretized counterpart of $u$ and $u^N_m$ is the approximation of $u^N$ on an $m$-dimensional Krylov subspace.  Finally, $\| \cdot \|$ denotes the Euclidean norm. 
   
We will assume that the square root function $\sqrt{z}$ (or equivalently $z^{1/2}$) has a branch cut on $(-\infty,0)$ and for $z\in \CC\setminus (-\infty,0)$ we assign the principal square root, i.e., $\Re \sqrt{z}>0$. 

Likewise, we assume the principal square root of a matrix or an operator,
i.e., for an operator $B$ with its spectrum on $\CC\setminus (-\infty,0)$ we assume that the spectrum of $\sqrt{B}$ (or equivalently $B^{1/2}$) lies in the open right half plane.

For $z\in (-\infty,0)$ we denote by $\sqrt{z\pm \text{i}0 }$ the corresponding limits of the principle values on the branch cut,
i.e.,$\sqrt{z\pm \text{i}0 }=\pm\text{i}\sqrt{-z}$.

%
%
\section{Problem formulation and necessary background}
\label{sec:stabcor}

\subsection{Self-adjoint formulation}
\label{subsec:selfadjoint}
To fix the idea, let us consider the scalar isotropic wave equation 
\begin{equation}
\label{eq:swec}
A \mathsf{u}-\mathsf{u}_{tt}=qb, 
\end{equation}
on $\RR^2\times[-\infty,\infty]$ with $q|_{t\le 0}=0$, $\mathsf{u}|_{t\le 0}=0$ and 
\[
A v=\frac{1}{c}\sum_{i=1}^2\left( v_{x_i}\right)_{x_i}
\] 
for any $v\in H^1[\RR^2]$. Here, $q$ is a real function of $t\in \RR$ and $b,c$ are real functions of $x=(x_1,x_2)\in \RR^2$. Furthermore, $q$ has bounded support, $b$ is supported in the square domain $\Omega=[-1,1]\times[-1,1]$,  $c>0$ in $\Omega$ and $c=1$ on $\Omega$'s complement. The functions~$b$ and $c$ satisfy the additional regularity conditions $b\in H^1[\RR^2]$ and $c\in L_\infty[\RR^2]$, and $q\in L_2[-\infty,\infty]$. Operator $A$ is $c$-self-adjoint (in the inner product with weight $c$) and negative definite with an absolutely continuous spectrum on $(-\infty,0)$, e.g., see~\cite{Taylor}. We note that the approach of this work is valid for any elliptic (or elliptic system) partial-differential operator $A$ with the same spectral properties. 

Unless specified otherwise, we shall assume that the pulse excitation is given by
\[
q(t)=\delta(t).
\]
In this case, problem~(\ref{eq:swec}) can be 
equivalently transformed to an initial-value problem on $\RR^2\times[0,\infty]$
\begin{equation}
\label{eq:hom}
A \mathsf{u}-\mathsf{u}_{tt}=0, \qquad \mathsf{u}|_{t=0}=0,\  \frac{\partial}{\partial t}\mathsf{u}|_{t=0}=-b,
\end{equation}
and the solution of (\ref{eq:swec}) and (\ref{eq:hom}) can be written in terms of operator functions as 
\begin{equation}
\label{eq:Fcosspec} 
\mathsf{u}=-\eta(t)(-A)^{-1/2} \sin \left(\sqrt{-A}t\right)b,
\end{equation} 
where $\eta(t)$ is the Heaviside unit-step function (e.g., see \cite{Taylor}).

By Laplace transforming (\ref{eq:swec}), we obtain the absorptive Helmholtz equation 
\begin{equation}
\label{eq:helm}
Au-\lambda u=b, \quad \lim_{\|x\|\to \infty} u=0,
\end{equation}
where  
\[u(\lambda)={\cal L}(\mathsf{u})(\lambda)=
\int_{-\infty}^{\infty} \exp(-t\sqrt{\lambda})\mathsf{u}(t)dt,
\]
is the two-sided Laplace transform of $\mathsf{u}$, which is well-defined for the principal value of $ \sqrt{\lambda}$.
The solution of equation~(\ref{eq:helm}) is obviously given by 
\begin{equation}
\label{eq:solLaplace} 
u=(A-\lambda I)^{-1}b
\end{equation}
and the resolvent $(A-\lambda I)^{-1}$ is an analytic function on $\CC\setminus (-\infty,0)$ with a branch cut on $(-\infty,0)$~\cite{Taylor}, i.e., the branch cut of $u$ coincides with the one of $\sqrt{\lambda}$. On the branch cut, we define $u$ via the limiting absorption principle  (e.g., see \cite{Taylor}), i.e.,
\[
u(\lambda\pm \text{i}0)=\lim_{\epsilon\to 0} u(\lambda\pm \text{i}\epsilon)
\] 
for real negative $\lambda$ and real positive $\epsilon$. Obviously, $u(\lambda)=\overline{u(\overline {\lambda})}$ on $\CC$ including the limiting branch cut values. 
The solution $u$ at the cut corresponds to the standard frequency domain solution, i.e., for $\lambda\in (-\infty,0)+\text{i}0$ we have an outgoing wave solution, while for $\lambda\in (-\infty,0)-\text{i}0$ we have an incoming wave solution. In addition, if we define the Fourier transform by
\[
{\cal F}(w)(\omega)=
\int_{-\infty}^{\infty} e^{- \text{i}\omega t} w(t) dt,
\]
then we have ${\cal F}(\mathsf{u})(\sqrt{-\lambda})=u(\lambda)$ for $\lambda\in (-\infty,0)\pm \text{i}0$. In other words, at the branch cut the Laplace transform turns into the Fourier transform. 
  
It is well known that for the solution of hyperbolic problems, spatial operators cannot be effectively approximated via discretization schemes that lead to self-adjoint matrices, because such discretizations would yield a discrete spectrum on $(-\infty,0)$ and consequently would invalidate the limiting absorption principle. Moreover, spurious resonances and reflections would be created, leading to gross qualitative errors in the numerical solution.

\subsection{Moving the branch cut away from real negative axis}
\label{subsec:movbr}
The resolvent can be analytically continued to another Riemann sheet in the neighborhood of the branch cut \cite{Taylor}, i.e, the branch cut can be moved away from the negative real axis. With the help of Berenger's perfectly matched layer~(PML) \cite{Berenger}, we shall construct operators with resolvents analytic in the neighborhood of $(-\infty,0)$ yielding such continuations.
  
Following \cite{ChewWeedon}, we introduce a PML on $\Omega$'s complement with the help of the complex coordinate transformation
\[
d\tilde x_i=\frac{dx_i}{\chi(\tilde x_i,\sqrt{\lambda})}, \ i=1,2,
\]
where the stretching factor is given by
\begin{equation}
\label{eq:def_str} 
\chi(r,s)=
\alpha(r)+
\frac{\beta(r)}{s},  
\end{equation}
with $\alpha(r)\ge 0$, $\beta(r) >0$  for $|r|>1$ and $\alpha(r)=1$, $\beta(r)=0$ otherwise.  
This stretching transforms $A$ into  
\[
\tilde A(\sqrt{\lambda})v=
\frac{1}{c}\sum_{i=1}^2\frac{1}{\chi({\tilde x}_i,\sqrt{\lambda})}
\left(\frac{1}{\chi({\tilde x}_i,\sqrt{\lambda})} v_{{\tilde x}_i}\right)_{{\tilde x}_i}.
\] 
The transformed operator is complex symmetric with respect to a weighted pseudo-inner product with a weight that is equal to the Jacobian's determinant times $c$, i.e., 
the weight is given by $c\prod_{i=1}^2\chi(\tilde x_i,\sqrt{\lambda})$.

Perfectly matched layers were originally introduced for the efficient truncation of unbounded computational domains \cite{Berenger}, so a plane wave Helmholtz solution $e^{-\sqrt{\lambda}k\cdot x}$ with $k=(k_1,k_2)$, $\|k\|=1$ and $\lambda\in (-\infty,0)$ becomes a decaying solution~\cite{ChewWeedon}, i.e., we have  
\[ 
\exp\left\{-\sum_{i=1}^2k_i\left[\text{i}\sqrt{-\lambda}\int\alpha({\tilde x}_i)d{\tilde x}_i +\int\beta({\tilde x}_i)d{\tilde x}_i\right]\right\}
\] 
in $\Omega$'s complement with an exponential decay rate given by $\beta$. Thanks to this decay, the Helmholtz equation 
\begin{equation}
\label{eq:PML} 
\tilde A(\sqrt{\lambda})  u'-\lambda  u'=b,\qquad
\lim_{\|\tilde x\|\to \infty}  u'=0
\end{equation}
has a unique solution $u'(\lambda,\tilde x)$ for $\Im \Lambda\ge 0$ assuming the positive branch of $\sqrt{\lambda}$. Moreover,  
\[
u'(\lambda, \tilde x)=u (\lambda, x)
\]
for $\tilde x=x\in \Omega$  and $\Im \Lambda\ge 0$, assuming both $u'$ and $u$ are on the same (positive) branch of $\sqrt{\lambda}$ (e.g., see \cite{Joly, ChewWeedon,Olyslager}).
The same result is valid for $\Im \Lambda\le 0$, assuming a negative branch of $\sqrt{\lambda}$, and $u'(\lambda, \tilde x)=\overline {u'(\overline \lambda, \tilde x)}$, similarly to $u$.

Let us assume that (\ref{eq:PML}) is approximated using a proper discretization scheme (e.g., Yee's algorithm) with $N$ nodes that preserves the weighted symmetry of $\tilde A(\sqrt{\lambda})$. We denote the state-vector, the right hand side vector, and the operator of the discretized problem by $ {u^N}'(\lambda)$, $b^N\in \RR^{N}$, and ${\tilde A}_N(\sqrt{\lambda})\in \RR^{N\times N}$, respectively. Then we can write $ {u^N}'(\lambda)$ as
\begin{equation}
\label{PMLresolvent}
{u^N}'(\lambda)=({\tilde A}_N(\sqrt{\lambda})-\lambda I)^{-1}b^N.
\end{equation}
Proper discretization yields a state vector ${u^N}'(\lambda)$ which is analytic on $\CC\setminus(-\infty,0)$ with a branch cut $(-\infty,0)$ and ${u^N}'$ converges to $u'$ on the entire complex plane including the corresponding limits on the branch cut (see, e.g., \cite{Joly}), i.e., the discretized problem preserves continuity of the spectral measure of  the original problem \cite{Olyslager}.
The dependence of ${\tilde A}_N(\sqrt{\lambda})$ on the spectral parameter $\lambda$ creates a nonlinear eigenproblem.
In principle, a ROM of (\ref{PMLresolvent}) can be designed with the help of an interpolatory projection (a.k.a. parameter-dependent Krylov subspace) method \cite{BG2008,dz}, but a drawback of such an approach is that it requires a full Arnoldi-type orthgonalization procedure. This can become exeedingly expensive for large~$N$ as is often required for the accurate spatial discretization of large scale wave problems. 

Now let us choose some $\lambda\in (-\infty,0)+\text{i}0$ and let $\omega_0=\sqrt{-\lambda}$ denote a fixed frequency that corresponds to this value of $\lambda$.  We consider the fixed frequency PML (FFPML) formulation of \cite{Bindel, Hislop&Sigal}:
\begin{equation}
\label{eq:PMLfix} 
\tilde A(\text{i}\omega_0) \tilde u-\lambda \tilde u=b,\qquad
\lim_{\|x\|\to \infty} \tilde u=0.
\end{equation}
First of all, we notice that for any imaginary $\sqrt{\lambda}$ of the same sign as $\omega_0$, equation (\ref{eq:PMLfix}) coincides with  (\ref{eq:PML}) using $\frac{\omega_0}{\sqrt{-\lambda}}\beta>0$ instead of $\beta$. Consequently, (\ref{eq:PMLfix})
has a unique solution $\tilde u$ that coincides with $u|_{\lambda\in (-\infty,0)+\text{i}0}$ for $\lambda<0$ and $x=\tilde x\in \Omega$.
The solution of the corresponding discretized fixed frequency formulation can be written as
\begin{equation}
\label{FFPMLresolvent}
\tilde u^N(\lambda) = (\tilde A_N -\lambda I)^{-1} b^N
\end{equation}
on the real negative semi-axis \cite{Bindel,Kim&Pasciak}, where we have written $\tilde A_N$ instead of $\tilde A_N(\text{i}\omega_0)$. We shall continue to use this notation further. Since the formulation of equation~(\ref{FFPMLresolvent}) corresponds to the linear non-Hermitian spectral problem
\[
\tilde A_N v_i -\lambda_i v_i=0,
\]
it is clear that FFPML linearizes the eigenproblem. It was originally intended to compute the resolvent and spectrum in some neighborhood of $-\omega_0^2$  \cite{Bindel, Kim&Pasciak}. 
  
\section{Discretization of the domain with the FFPML}
\label{sec:discretization}

\subsection{ Discrete FFPML via optimal rational approximation of the square root}
\label{subsec:discrFFPML}

Our objective is to obtain an efficient spatial discretization in $\Omega$'s complement for a given frequency range. The drawback of FFPML is that, unlike Berenger's PML, it does not scale the attenuation factor (or the imaginary part of the grid coordinate) with the wavelength. The quality of the approximations may therefore deteriorate away from $\omega_0$. 

This drawback can be circumvented, however, by adopting the optimal grid approach of \cite{AsDrKn1, Ingerman_etal} (a.k.a. finite-difference Gaussian quadratures or spectrally matched grids). This approach allows us to design a discrete FFPML via optimal rational approximations on a given spectral range. 

To explain how this is realized, let us consider an FFPML that occupies the half-plane ${\tilde x}_1\ge 0$ of $\RR^2$ with boundary at $\tilde{x}_1=0$. We consider the equation
\begin{equation}
\label{eq:swe}
\sum_{i=1}^2\frac{\partial^2 \mathsf{u}}{\partial {\tilde x}_i^2}-\mathsf{u}_{tt}=0
\end{equation} 
for ${\tilde x}_1\ge 0$.
Applying a spatial Fourier transform with respect to the ${\tilde x}_2$-coordinate and the Laplace transform with respect to $t$, we obtain
\begin{equation}
\label{zero}
w_{rr} - s w =0,
\end{equation}
with 
\begin{equation}
s = \kappa^{2} +\lambda,
\end{equation}
and where $\lambda$ is the complex Laplace parameter as defined in Section~\ref{sec:stabcor}. Furthermore, $\kappa$ is the spatial Fourier frequency and we have slightly abused notation by denoting $r=\tilde{x}_1$.

Equation (\ref{zero}) has two solutions $\exp\mp\sqrt{s}t$,
and we are interested in the outgoing one given by $\exp-\sqrt{s}t$  for  $\lambda\in (-\infty,
0)+\text{i}0$, which is obtained from the limiting absorption principle.

Scaled outgoing solutions are defined by their  Neumann to Dirichlet map (NtD) at the FFPML boundary $r=0$,
i.e., 
\begin{equation}
\label{dtn}\frac{w(0)}{w(0)_{r}}=-\frac{1}{\sqrt{s}}.
\end{equation}

We assume that the Fourier spectrum of the time-domain solution  
is supported on the positive frequency interval [$\omega_{\text{min}},\omega_{\text{max}}]$. Consequently, we have that $\lambda\in (-\omega_{\text{max}}^{2},
-\omega_{\text{min}}^2)+\text{i}0$. 

We will be interested only in propagating waves for which $\kappa^{2}< \omega^2$. The discussed approach can be extended to evanescent waves ($\kappa^{2}\ge \omega^2$, see~\cite {DGH}), but these waves can also be handled at some insignificant cost by simply distancing the FFPML boundary from the actual domain of interest. 
In addition, for propagating waves we set $\mu=[1-(\frac{\kappa}{\omega})^{2}]^{1/2}$ and we may bound the range of incidence angles of these waves by imposing the constraint $\mu^2\le 1-(\frac{\kappa}{\omega})^2$. The approach considered here allows for the inclusion of the case $\mu=0$~\cite{AsDrKn1}, but for simplicity we shall take $\mu>0$.
Putting everything together, it follows that the interval of interest for $s$ is given by $[s_{\text{min}},s_{\text{max}}]$ with 
\begin{equation}
\label{sinterval} 
s_{\text{min}}=-\omega_{\text{max}}^2
\quad \text{and} \quad 
s_{\text{max}}=-(\omega_{\text{min}}\mu)^2.
\end{equation}
To summarize, we want to obtain a discrete system that approximates $1/\sqrt{s}$
on the interval $[s_{\text{min}},s_{\text{max}}]+\text{i}0$. To this end, we restrict the approximant to an analytic function of $s$ on $\CC\setminus(0,\infty)$, i.e., we move the branch cut from the negative to the positive real semi-axis and consider an approximation of the main branch of the square root. Further, from now on we omit the branch cut limit notation $+\text{i}0$ in this section.

For the case $\omega_{\text{min}}=\omega_{\text{max}}=1$, the spectral interval $[s_{\text{min}},s_{\text{max}}]$ coincides with the one considered in \cite{AsDrKn1}, so we shall just follow their derivation using the modified spectral interval.

Let us approximate the solution $u$ to (\ref{zero}) 
by a staggered three-point finite difference scheme.
In a staggered scheme, the numerical solution is defined at ``potential'' (primary) nodes
${r}_i$, $i=1,...,k+1$,   
with ${r}_1=0$, and the finite difference 
derivatives are defined at ``derivative'' (dual) nodes 
$\hat r_i$, $i=0,\dots,k$, with $\hat{r}_0=0$. 
We  denote the {\it complex} step sizes by 
$h_i=r_{i+1}-r_{i}$ and $\hat h_i=\hat r_{i}-\hat r_{i-1}$, respectively,  
and solve the following finite difference problem
\begin{equation} 
\frac{1}{\hat h_i}\left(\frac{w_{i+1}-w_i}{h_{i}}-
\frac{w_{i}-w_{i-1}}{h_{i-1}}\right)-s w_i=
0, \qquad i=2,\ldots,k,
\label{fdd} 
\end{equation}
with boundary conditions
\[
\frac{1}{\hat h_1}\left(\frac{w_{2}-w_1}{h_{1}}\right)-s w_1=
-\frac{1}{\hat h_1}$$ and $$ \qquad w_{k+1}=0.
\]
Note that the first boundary condition 
is consistent with the differential equation since it is
the same as creating a dummy node $w_0$, allowing $i=1$ in
(\ref{fdd}) and setting
\[
\frac{w_{1}-w_{0}}{h_{0}}=-1.
\]
We express the linear system (\ref{fdd}) for $w$ in shorthand by
$(L_h-s)w=-\frac{1}{\hat h_1}e_1 $,
where $e_1$ is the unit vector with support in the first component. 
The continuous, or true, impedance function $\varphi(s)$ is defined 
by $$\varphi(s)\equiv w(0),$$ and the
discrete, or approximate,  impedance function $\varphi_k(s)$ is defined by  
$$\varphi_k(s)\equiv w_1.$$ 

Our objective is to choose the placement of the grid points such that the
discrete impedance function $\varphi_k(s)$ is an accurate approximation
to $1/\sqrt{s}$ on the interval of interest $ [s_{\text{min}},s_{\text{max}}]$.

Suppose we approximate the solution to (\ref{zero}) by the finite 
difference solution to (\ref{fdd}). Then
we can represent the discrete impedance in terms of the eigenpairs of
the matrix $L_h$. Note that $L_h$ is not symmetric in the standard sense,
but is (complex) symmetric in the pseudo-inner product with weights ${\hat h}_i$,
$$\langle x,y\rangle_{\hat h}=\sum_{i=1}^k {\hat h}_ix_iy_i,$$
that is,
$$ \langle L_h x,y\rangle_{\hat h}= \langle x,L_h y\rangle_{\hat h}$$
for any $x,y\in R^k$.  
  
 Let $z_i,\theta_i$ be the eigenvectors and eigenvalues
(respectively) of the matrix $L_h$, normalized with respect to the inner
product $\langle ,\rangle_{\hat h}$. 
Then the discrete impedance function can be written as
\begin{equation} 
\label{disimp} 
\varphi_k(s)= \sum_{i=1}^k\frac{y_i}{s-\theta_i}  
\quad \text{where} \quad
y_i=(z_i)^2_1.
\end{equation}
This shows that $\varphi_k$ is a $[k-1/k]$ rational function.  We will find it as the best (relative) real $[k-1/k]$ rational approximation of $1/\sqrt{s}$ on $[s_{\text{min}},s_{\text{max}}]$, i.e., by minimizing
\[\varphi_k=
\underset{\theta_1,\ldots,\theta_k,y_1,\ldots,y_k}{\text{argmin}}\left[\max_{s\in[s_{\text{min}},s_{\text{max}}]}\left|1- \sqrt{s}\varphi_k(s)\right|\right] \]
with $\varphi_k$ in the form (\ref{disimp}) with real $y_{i}$ and $\theta_{i}$. 
The explicit optimal solution of this problem was obtained by the Russian mathematician E.~Zolotarev in 1877, (see, e.g., \cite{PetrPop}) and its parameters can be computed via elliptic integrals. 

The Zolotarev solution yields real (noncoinciding) negative poles $\theta_i$ and real positive residues $y_i$ and as such can be  uniquely represented in the form of a Stieltjes continued fraction 
\begin{equation} \label{fraction}
\varphi_k(s) =
 \cfrac{1}{\hat h_1\lambda+
 \cfrac{1}{h_1+
 \cfrac{1}{\hat h_2\lambda+\dots
 \cfrac{1}{h_{k-1}+
 \cfrac{1}{\hat h_{k}s+
 \cfrac{1}{h_k}}}}}}
\end{equation}
with pure imaginary coefficients $\hat h_l=\text{i}\hat \gamma_l$,
$ h_l=\text{i} \gamma_l$, $\hat \gamma_l,\gamma_l>0$.
Thus, (\ref{fdd}) can be considered as a finite-difference discretization of the FFPML with pure imaginary stretching. Its coefficients depend on the approximation interval $[s_{\text{min}},s_{\text{max}}]$ via the Zolotarev solution. In this way, we have avoided explicit dependence on the fixed frequency $\omega_0$.

Given an impedance function of the form (\ref{disimp}), the step sizes $h_i$ and $\hat h_i$ can be obtained from the parameters $y_i$ and $\theta_i$ by 
equating (\ref{disimp}) to (\ref{fraction})
and using the Euclidean polynomial division algorithm via the
Lanczos algorithm~\cite{Druskin&Knizhnerman}. We now have the optimal step sizes available. As was mentioned in the introduction, these step sizes are computed a priori and only once. 

\begin{figure}
\begin{center}
\includegraphics*[width=0.95\textwidth]{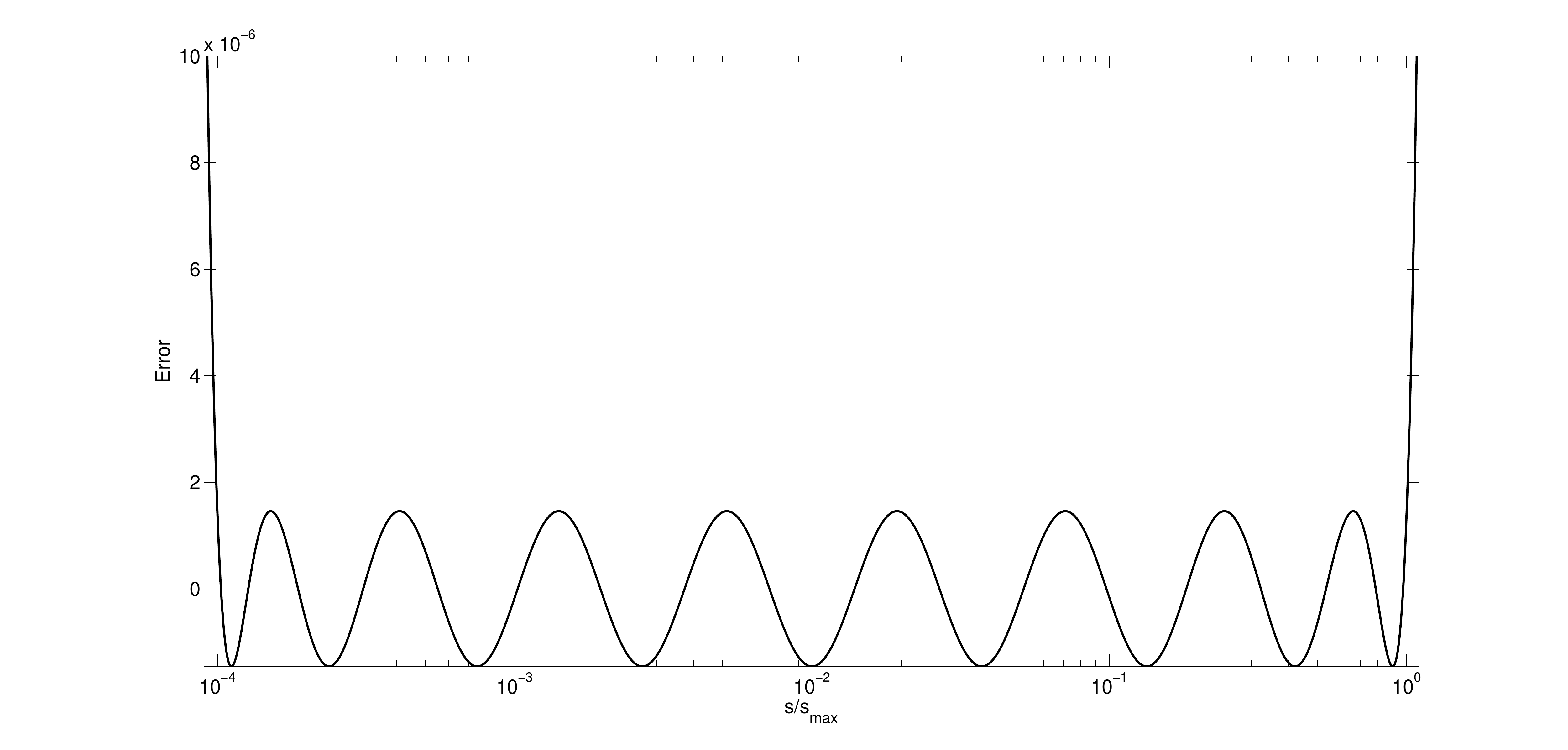} \\
\includegraphics*[width=0.95\textwidth]{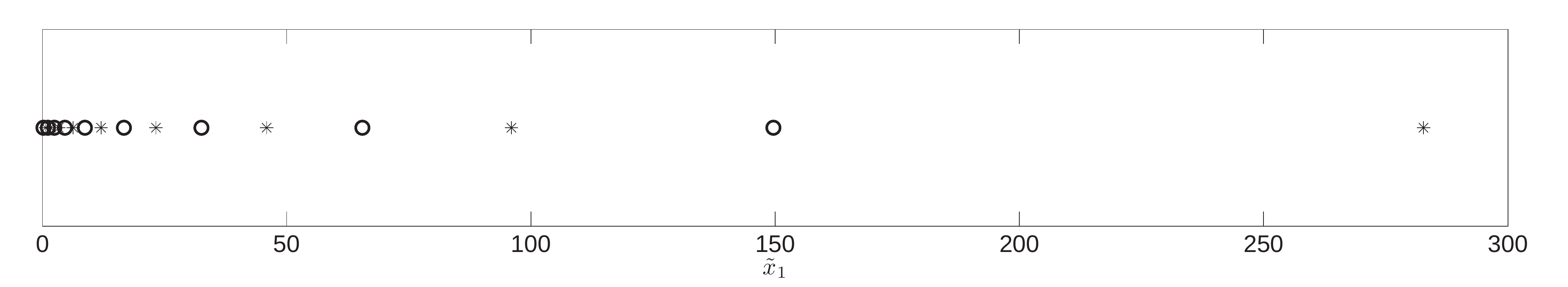}
\end{center}
\caption{Impedance error $1- \sqrt{s}\varphi_k(s)$ for an optimal Zolotarev grid with $\chi=10^{4}$ and $k=9$ (top) and the corresponding optimal grid nodes~(bottom). The optimization interval is~$[10^{-4},1]$. The crosses and circles in the bottom figure indicate the location of the primary and dual  nodes, respectively. Notice that the grid nodes cluster near the inner FFPML boundary.}
\label{fig:impedance_error&grid}
\end{figure}

The beauty of the optimal Zolotarev solution is that it has exponential convergence with a rate that is very weakly dependent on the interval condition number $\chi=\frac{s_{\text{max}}}{s_{\text{min}}}$. Specifically, for large enough $\chi$ the optimal error behaves as~\cite{Ingerman_etal}
\begin{equation}
\label{zolest}
\max_{s\in[s_{\text{min}},s_{\text{max}}]}\left|1-\sqrt{s}\varphi_k(s)\right| =
O\left(e^{-\frac{\pi^2[1+o(1)]}{2\log
\chi}k}\right), 
\end{equation}
so with a rather small number of finite-difference nodes one can obtain a very accurate FFPML discretization on large frequency intervals and wide ranges of incidence angles.
As an illustration, Figure~\ref{fig:impedance_error&grid}~(top) shows the Zolotarev impedance error for a condition number $\chi=10^{4}$ and $k=9$. The maximum absolute error on the optimization interval~$[10^{-4},1]$ is $1.46 \cdot 10^{-6}$. Also note the dramatic increase of the error just outside this interval. The corresponding optimal grid nodes are shown in Figure~\ref{fig:impedance_error&grid}~(bottom). The grid is aligned with the imaginary axis and refines towards the left end, which is the inner FFPML boundary.

\subsection{Discretization of the entire computational domain}
\label{subsec:discr_entire_domain}

We will use the optimal (Zolotarev) finite-difference scheme described above for the discretization of the complement of $\Omega$. Since it has spectral (exponential) accuracy given by (\ref{zolest}), it is preferable to discretize $\Omega$ using an algorithm with consistent accuracy, i.e., a high-order spectral method (see, e.g., \cite{Hesthaven_etal}) or, alternatively, the optimal grid approach for interior domains (see, e.g., \cite{AsDrKn1}).  However, for simplicity, we shall use an equidistant second order scheme in the interior and obtain the standard five-point finite-difference scheme throughout the entire computational domain.

First, let us consider the primary nodes. These nodes are given by $ d_i$, $0\le i\le 2k+1+n$, with $d_0=-1-r_{k+1}$ and $r_{2k+1+n}=1+r_{k+1}$. The primary step sizes are  $d_{i}-d_{i-1}=h_{k+1-i}$ for $i=1,\ldots,k$, $d_{i}-d_{i-1}=\frac{1}{n}$ for $i=k+1,\ldots,k+n+1$, and $d_{i}-d_{i-1}=h_{i-k-n}$ for $i=k+n+1,\ldots,2k+n+1$. Second, the dual nodes are given by $\hat d_i$, $1\le i\le 2k+n$, with $\hat d_1=-1-\hat r_k$ and $d_{2k+n}=1+\hat d_k$. The dual step sizes are $\hat d_{i}-\hat d_{i-1}=\hat h_{k-i}$ for $i=2,\ldots,k-1$,  $\hat d_{k}-\hat d_{k-1}=\hat h_1+\frac{1}{2n}$,  $\hat d_{i}-\hat d_{i-1}=\frac{1}{n}$ for $i=k+1,\ldots,k+n-1$, and $\hat d_{k+n}-\hat d_{k+n-1}=\hat h_1+\frac{1}{2n}$ and $\hat d_{i}-\hat d_{i-1}=\hat h_{i-k-n}$ for $i=k+n+1,\ldots,2k+n$. 

We define $\tilde u^N$ on the two-dimensional primary grid $G_N$ with nodes
$x_{{i_1},{i_2}}=(d_{i_1},d_{i_2})$, $N=(2k+n)^2$, and obtain the finite-difference form of (\ref{FFPMLresolvent}) for $0< i_1,i_2<2k+1+n$ as 
\begin{eqnarray}
\label{full2D}
\frac{1} {\hat d_{i_1}-\hat d_{i_1-1}}\left(\frac{{\tilde u^N}_{i_1+1,i_2}-{\tilde u^N}_{i_1,i_2}}{ d_{i_1+1}-d_{i_1}}
-\frac{{\tilde u^N}_{i_1,i_2}-{\tilde u^N}_{i_1-1,i_2}}{ d_{i_1}-d_{i_1-1}}\right)&+&\\
\frac{1} {\hat d_{i_2}-\hat d_{i_2-1}}\left(\frac{{\tilde u^N}_{i_1,i_2+1}-{\tilde u^N}_{i_1,i_2}}{ d_{i_2+1}-d_{2_1}}
-\frac{{\tilde u^N}_{i_1,i_2}-{\tilde u^N}_{i_1,i_2-1}}{ d_{i_2}-d_{i_2-1}}\right)&-& \lambda c^N_{i_1,i_2}{\tilde u^N}_{i_1,i_2}=
{b^N}_{i_1,i_2},\nonumber\\
{\tilde u^N}_{0,i_2}={\tilde u^N}_{2k+1+n,i_2}={\tilde u^N}_{i_1,0}={\tilde u^N}_{i_1,2k+1+n}&=&0,\nonumber
\end{eqnarray}
where $c^N_{i_1,i_2}$ and ${b^N}_{i_1,i_2}$ are the nodal values of $c$ and $b$, respectively.

Equation (\ref{full2D}) explicitly defines $\tilde A_N$. It is complex symmetric matrix with respect to a pseudo-inner product with weight $(\hat d_{i_1}-\hat d_{i_1-1})(\hat d_{i_2}-\hat d_{i_2-1})c^N_{i_1,i_2}$, i.e., 
\[M\TAN=\TAN^TM\]
($M$-symmetric), where 
\[
M=\text{diag} 
\left[(\hat d_{i_1}-\hat d_{i_1-1})(\hat d_{i_2}-\hat d_{i_2-1})c^N_{i_1,i_2}
\right].
\]
For regular enough $b$ and $c$ the finite-difference truncation error of the uniform grid in $\Omega$ will be $O(n^{-2})$,  so the total error of (\ref{full2D}) can be estimated as 
\begin{eqnarray}
\label{toterr}
\left|u(d_{i_1},d_{i_1})-{\tilde u^N}_{i_1,i_2}\right|= 
O(n^{-2})+O\left(\max_{s\in[s_{\text{min}},s_{\text{max}}]}
\left|1-\sqrt{s}\varphi_k(s))\right|\right)\approx \\ O(n^{-2})+O\left(e^{-\frac{\pi^2[1+o(1)]}{2\log
\chi}k}\right),
\nonumber
\end{eqnarray}
assuming that $x=(d_{i_1},d_{i_1})\in \Omega$ and  $\lambda$, $\mu$ satisfy (\ref{sinterval}) \cite{Asvadurov_etal}. 

\section{Stable time-domain solution via a damped operator function}
\label{SCTDE}
\subsection{Stability-corrected time-domain exponent}
Convergence  of $\tilde u^N=(\TAN -\lambda I)^{-1} b^N$ on the real negative semi-axis is not sufficient for convergence on the entire complex plane.
The spectrum of the complex non-Hermitian matrix ${\tilde A_N}$ is moved from the real negative semi-axis, i.e.,  (unlike $u$ and $u'$) $\tilde u^N$ has poles on $\CC\setminus (-\infty,0)$, and therefore $\tilde u^N(\lambda)$ loses convergence away from the real negative semi-axis. Moreover, a straightforward inverse Fourier transform of (\ref{FFPMLresolvent}) to the time domain would yield  for $t>0$ a representation with the same operator function as in (\ref{eq:Fcosspec}),
but with $\TAN$ instead of  $A$. Due to the presence of a nontrivial imaginary part in ${\tilde A}_N$'s spectrum,  $\sin(\sqrt{\TAN}t)$ would grow exponentially with $t$.

To circumvent these problems, we define the stability corrected time-domain exponent (SCTDE) for an impulse excitation as
\begin{equation}
\label{SKexp} 
\mathsf{u}^N=\eta(t)\Re \mathsf{f} (t,\tilde A_N)b^N 
\quad \text{with} \quad 
\mathsf {f}(t,a)= \left[\frac{e^{ -\sqrt{a}t}}{\sqrt{a}}\right].
\end{equation}
If $\TAN$ is diagonalizable, i.e., there exist $\lambda_i\in\CC$  and $v_i\in\CC^N$, with $v_i^T Mv_j=\delta _{i,j}$ ($\delta_{i,j}$ is the Kronecker delta function), such that
\[\TAN v_i=\lambda_i v_i, \quad i=1,\ldots,N,\]
then (\ref{SKexp}) can be {\it formally} represented via a spectral decomposition
as
\begin{equation}
\label{SD} 
\mathsf{u}^N=\eta(t)\Re \sum_{i=1}^Nv_i\mathsf{f} (t,\tilde \lambda_i)(v_i^TMb^N). 
\end{equation}

Having the impulse response available, the solution for a general excitation~$q$ can be obtained via a temporal convolution, i.e.,
\[
\mathsf{u}^N= \Re \mathsf{f}_{\text{q}} (t,\tilde A_N)b^N 
\quad \text{with} \quad 
\mathsf {f}_{\text{q}}(t,a)=\int_{-\infty}^\infty \eta(t-\tau)\frac{e^{ -\sqrt{a}(t-\tau)}}{\sqrt{a}}q(\tau)
\ d \tau.
\]

Observe that equation~(\ref{SKexp}) becomes identical to (\ref{eq:Fcosspec}) if we replace $\TAN$ by $A$. However, according to the principal square root convention given in subsection~\ref{subsec:convention}, (\ref{SD}) does not contain any growing exponents for non-Hermitian $\TAN$. Intuitively, the SCTDE can be understood by reasoning that, physically, the solution should be a symmetric function of  $\tilde A_N(\text {i}\omega_{0})$ and $\tilde A_N(-\text {i}\omega_{0})=\overline {\tilde A_N(\text {i}\omega_{0})}$, corresponding  to FFPMLs at both the sides of the branch cut.  A rigorous justification of the SCTDE is given in the following subsection.

\subsection{Plancherel's identity for the SCTDE error}
\label{sec:Plancherel}

In this section, we derive a Plancherel-like identity connecting the $L_2$ norm of $u^N-u$ on the imaginary positive semiaxis axis and the $L_2$ norm of $\mathsf{u}^{N} -\mathsf{u}$ on the real positive semiaxis. It  implies, that if the $L_2$ frequency domain discretization error of $u^N$ vanishes in $\Omega$ as $N\rightarrow \infty$, then the $L_2$ time domain SCTDE error of $\mathsf{u}^{N}$ vanishes as well. 

We start by considering the Laplace transform $\hat u^N$ of $\mathsf{u}^{N}$. This transformed field quantity can be written as   
$\hat u^N=\mathcal{L}(\mathsf{u}^{N})(\lambda)= f[\lambda,{\tilde A}_N]b^N$, where
\begin{equation}
\label{SKR}
f(\lambda,\TAN)=\frac{1}{2}\TAN^{-1/2}\left(\sqrt{\lambda}I+\sqrt{\TAN}\right)^{-1}+\frac{1}{2}\overline{\TAN^{-1/2}}\left(\sqrt{\lambda}I  +\overline{\sqrt{ \TAN}}\right)^{-1}.
\end{equation}
If $\TAN$ has its spectrum outside $(-\infty,0)$, the function $f(\lambda,\TAN)$ is analytic with respect to $\lambda$ in  $ \CC\setminus (-\infty,0)$ with $(-\infty,0)$ being the branch cut. Obviously, $\hat u^N$ (as a function of $\lambda$) inherits these properties and we recall that for any fixed $x$ the exact frequency domain solution $u$ also has these analytic properties as a function of $\lambda$. 

\begin{lemma}\label{SCRF}
Let us assume that  $\tilde A_N$ has its spectrum outside  $(-\infty,0)$.
Then  $\forall\lambda \in (-\infty,0)+0\text{i}$, we have $\Re\hat u^N=\Re\tilde u^N$.
\end{lemma}

\bigskip
\noindent
{\it Proof.\ \ \ }  
With 
\begin{align*}
r(\lambda,\TAN) &= f(\lambda,\TAN) -\left(\TAN-\lambda I\right)^{-1} \\
&=\frac{1}{2}\TAN^{-1/2}\left(\sqrt{\lambda}I-\sqrt{\TAN}\right)^{-1}+\frac{1}{2}\overline{\TAN^{-1/2}}\left(\sqrt{\lambda}I  +\overline{\sqrt{ \TAN}}\right)^{-1},
\end{align*}
and for $\lambda\in (-\infty,0)+0\text{i}$,  the obvious identity 
\[r(\lambda,\TAN)\equiv -\overline {r(\lambda,\TAN)}\]
 yields us 
 \[
 \Re r(\lambda,\TAN)\equiv 0
 \] 
for the same value of $\lambda$. $\square$
 
\bigskip
\noindent 
To present our Plancherel identity for the SCTDE error, we will need the following known modification of Plancherel's theorem~\cite{Yosida}.
\begin{lemma}
\label{lem0}
Let $w(t) \in L_2(-\infty,\infty)\cap L_1(-\infty,\infty)$ and $w(t)=0$ for $t<0$. Then \[\int_{-\infty}^\infty w^2 dt=\frac{2}{\pi}\int_{0}^\infty \left[\Re {\cal F} (w)\right]^2 d\omega .\]
\end{lemma}

\bigskip
\noindent
 {\it Proof.\ \ \ } 
First we notice that $\int_{-\infty}^\infty w^2 dt=2\int_{-\infty}^{\infty} w'(t)^2dt$, where 
\[
w'(t)=0.5\left[w(t)+w(-t)\right].\]
Due to the regularity assumption on $w$, namely, $w(t)\in L_2(-\infty,\infty)\cap L_1(-\infty,\infty)$, we can apply Plancherel's identity $\int_{-\infty}^{\infty} {w'}^2dt=\frac{1}{2\pi}\int_{-\infty}^{\infty}  |{\cal F}(w')|^2 d\omega$ and obtain
$\int_{-\infty}^\infty w^2 dt=\frac{1}{\pi}\int_{-\infty}^{\infty}  |{\cal F}(w')|^2 d\omega$.
By construction, $w'(t)$ is a real and even function of $t$ and therefore ${\cal F}(w')=\Re {\cal F}(w)$ is the cosine transform of $w$, and as such it is a real and even function of $\omega$. With this result, we obtain 
\[
\int_{-\infty}^\infty w^2 dt
=\frac{1}{\pi}\int_{-\infty}^{\infty} [\Re {\cal F}(w)]^2 d\omega=\frac{2}{\pi}\int_{0}^{\infty} [\Re {\cal F}(w)]^2 d\omega.
\]
\hfill $\square$

We assume a regular enough $b$, so that the solution $\mathsf{u}$ as function of time is both from $L_2(-\infty,\infty)$ and $L_1(-\infty,\infty)$ \cite{Taylor}. The same is obviously true for $\mathsf{u}^N$ provided $\tilde A_N$'s spectrum is outside $(-\infty,0)$. 
 
We are now in a position to formulate our main (Plancherel-like) result relating the $L_2$ time-domain error of the SCTDE solution to the $L_2$ frequency domain error of the real part of the FFPML solution.
The approximate solutions $\mathsf{u}^N$ and $\hat u^N$ are normally defined at the nodes of the discretization scheme. Let $ x'\in \Omega$ be such a node, and define the time and frequency domain error functions at $x=\tilde x=x'$ as, respectively,
\begin{align*}
\delta_N(t)&=\mathsf{u}^N-\mathsf{u} \qquad \text{with $t \in \RR$},
\intertext{and}
\tilde \delta_N(\sqrt{\lambda} )&=\Re (u-\tilde u^N) \qquad \text{with $\lambda \in (-\infty,0)+\text{i}0$},
\end{align*}
i.e., $\arg \delta_N \in (0,\infty)$.

\begin{proposition}\label{main}
If the spectrum of $\tilde A_N$ belongs to $\CC\setminus (-\infty,0)$ and $\mathsf{u}|_{x=x'}$ is both from $L_2(-\infty,\infty)$ and $L_1(-\infty,\infty)$, then
\begin{equation}
\label{mainident}
\int_{-\infty}^{\infty} \delta_N^2dt
 =\frac{2}{\pi}\int_{0}^{ \infty}\tilde{\delta}_N^2 d\omega.
 \end{equation}
 \end{proposition}
 
 \bigskip
 \noindent
 {\it Proof.\ \ \ } 
 By definition, $\delta_N(t)=0$ for $t<0$ and $\delta_N(t) \in L_2(-\infty,\infty)\cap L_1(-\infty,\infty)$ due to the assumptions of the proposition on $\mathsf{u}^N$ and $\tilde A_N$. Consequently, 
\begin{equation}
\label{Plancherel0.5}
\int_{-\infty}^{\infty} \delta_N^2dt=\frac{2}{\pi}\int_{0}^\infty \left[\Re {\cal F} (\delta_N)\right]^2 d\omega,
 \end{equation}
since $\delta_N(t)$ satisfies the conditions of Lemma~\ref{lem0}. 
For $-\omega^2=\lambda\in (-\infty,0)+\text{i}0$, the Fourier and
Laplace transforms coincide, i.e., $\Re({\cal F} (\delta_N))=\Re(u-\hat u^N)$. Lemma~\ref{SCRF} allows us to replace $\hat u^N$ with $\tilde u^N$ in the last equality. Substituting this into  (\ref{Plancherel0.5}) we obtain (\ref{mainident}).~$\square$
 
The obtained result gives us the following justification for the SCTDE.
If $\tilde u^N$ converges to $u$ for some $x\in \Omega$ in the frequency domain $L_2$ norm, then  Proposition~\ref{main} yields convergence of $\mathsf{u}^N$ to $\mathsf{u}$ (for the same $x$) in the $L_2$ time-domain norm. Finally, we like to point out that all the results of this section are straightforwardly extended to any excitation $q(t)$ by introducing a weight $|{\cal F}(q)(\omega)|^2$ in the frequency domain error norm. 

\section{Krylov subspace projection algorithm}
\label{sec:krylov}

Let $U_m$ be an $m$-dimensional projection subspace of $\CC^N$ such that $m<<N$. We introduce a basis matrix~$V_m\in \CC^{N\times m}$ ($U_m=\hbox{colspace } V_m$) such that $V_m$ satisfies the quasi-M-orthogonality condition 
\[
V_m^T M V_m=I_m,
\] 
where $I_m$ is the $m\times m$ identity matrix. Let $H_m$ be the projection of $\TAN$ on $U_m$ given by 
\[
H_m=V_m^TM\TAN V_m.
\]
Then for any function $g(z)$, continuous on the spectra of both $\TAN$ and $H_m$, we can formally define the approximation 
\[ 
g(\TAN)b^N\approx V_m g(H_m)V_m^TM b^N.
\]
Such an approximation is efficient if it is accurate with $m<<N$.
In particular, we define the approximate solution $\mathsf{u}_m^N\approx \mathsf{u}^N$ as 
\begin{equation}
\label{subapprox} 
\mathsf{u}_m^N= \eta(t)\Re V_m\mathsf{f} (t,H_m)V_m^TMb^N.
\end{equation}
Due to the complex finite-difference steps, matrix~$M$ is complex symmetric and $\TAN$ is complex $M$-symmetric. Consequently, (\ref{subapprox}) is a Galerkin-Petrov approximation. It would be a Galerkin approximation if we used $\overline V_m^T$ instead of $V_m^T$ in the above formulas, or if $\TAN$ and $M$ were real with $M$ definite.
Unlike the Galerkin method, however, the Galerkin-Petrov method does not allow us to make any prediction about the spectrum of the projected matrix.
However, $\mathsf{f} (t,z)$ is a continuous function of $z$ on $\CC$ and for any $z\in \CC$, $\mathsf{f}(t,z)$ is a nonincreasing function of $t$ (actually, monotonically decreasing for $z\in \CC\setminus (-\infty,0)$). This implies that the approximations of (\ref{subapprox}) are always stable. 

As a projection subspace, we take the Krylov subspace generated by $\TAN$ and $b^{N}$, i.e., $U_m=\hbox{span}\{b^N,\TAN b^N,...,\TAN ^{m-1}b^N\}$. Since matrix~$\TAN$ is $M$-symmetric, a quasi-$M$-orthonormal basis can be efficiently constructed via the three-term bi-Lanczos recursion \cite{Freund} 
\begin{equation}
\label{biLanczos}
\beta_{i+1}v_{i+1}=\TAN v_i-\alpha_iv_i-\beta_iv_{i-1},\qquad i=1,\ldots,m,
\end{equation}
with initial data $v_{0}=0$, $\beta_{1}=((b^N)^T M b^N)^{1/2}$ and $v_{1}=\beta_{1}^{-1}b^{N}$. Here, the $v_i\in\CC^N$ and the 
recursion coefficients $\alpha_i \in \CC$ and $\beta_i \in \CC$ are obtained from the quasi-orthonormality conditions $v_{i+1}^TMv_i=0$ and 
$v_{i+1}^TMv_{i+1}=1$, respectively. This algorithm coincides with the classical Lanczos algorithm if $M$ and $A$ are real symmetric and $M$ is definite. 
The  three-term recursion not only gives an economical formula of computing
$V_m=(v_1,\ldots,v_m)$, it also yields a symmetric (complex) tridiagonal $H_m$ with main diagonal $\alpha_{1},\ldots,\alpha_m$ and subdiagonal(s) $\beta_{2},\ldots,\beta_{m}$. 
By construction $V_m^TMb^N= \beta_{1} e_1$, where $e_1$ is  the first column of $I_{m}$, so we can simplify (\ref{subapprox}) to
\begin{equation}
\label{subapproxL} 
\mathsf{u}_m^N= \beta_{1} \eta(t)\Re V_m\mathsf{f} (t,H_m)e_1.
\end{equation}

\begin{remark}It is well known that even the classical Lanczos recursion for real symmetric matrices is unstable in computer arithmetic, i.e., the Lanczos vectors lose global orthogonality. However, the classical Lanczos recursion (without reorthogonalization) still allows for the efficient approximation of matrix functions of the form (\ref{subapproxL}), with computer round-off just slightly affecting convergence of large scale problems~\cite{DGK}. For non-Hermitian matrices as considered here, the behavior of the Lanczos algorithm is significantly more complicated. In particular,
we observed significant growth of the true Euclidean norm $\|v_i\|$ (up to $10^7$), that affected the stability of (\ref{subapproxL}). To circumvent this problem,  we followed~\cite{Freund} and instead of (\ref{biLanczos}) used the algebraically equivalent but computationally more stable recursion in terms of normalized Lanczos vectors that have a Euclidean norm equal to one. Specifically, the basis vectors are generated via the recursion 
\[
\zeta_{i+1} w_{i+1} = \TAN w_{i} - \alpha_{i} w_{i} - \delta_{i} \delta_{i-1}^{-1} \zeta_{i} w_{i-1},
\]
with starting values $w_{0}=0$, $\delta_{0}=1$, $\zeta_{1}=\| b^N\|$, and $w_{1}=\zeta_{1}^{-1} b^N$. Furthermore, the coefficients $\zeta_{i}$ follow from the condition 
$\| w_{i} \|=1$, and the coefficients $\delta_{i}$ and $\alpha_{i}$ are given by $\delta_{i} = w_{i}^{T} M w_{i}$, and $\alpha_{i} = \delta_{i}^{-1} w_{i}^{T} M\TAN w_{i}$, respectively. 

After a successful completion of this algorithm, we have the Lanczos decomposition 
\begin{equation}
\label{eq:LancDec}
\TAN W_{m} = W_{m} T_{m} + \zeta_{m+1} w_{m+1} e_{m}^{T},
\end{equation}
where $e_{m}$ is the $m$th column of $I_{m}$ and the basis matrix $W_{m}=(w_{1},w_{2},...,w_{m})$ satisfies (in exact arithmetic)
\[
W_{m}^{T} M W_{m} = \text{diag}(\delta_{1},\delta_{2},...,\delta_{m})=:D_{m}.
\]
Furthermore, matrix~$T_{m}$ is a tridiagonal $m$-by-$m$ matrix containing the recurrence coefficients and is given by 
\[
T_{m} = \text{tridiag}(\zeta_{i}, \alpha_{i}, \delta_{i+1} \delta_{i}^{-1} \zeta_{i+1}).
\]
Notice that $V_{m}=W_{m} D_{m}^{-1/2}$ and $H_{m}$ is similar to $T_{m}$ with similarity matrix~$D_{m}^{-1/2}$, i.e., $H_{m} = D_{m}^{1/2} T_{m} D_{m}^{-1/2}$. 
  \end{remark}

The tridiagonal structure of $H_m$ allows for the efficient computation of the time-dependent vector $\mathsf{f} (t,H_m)e_1$. For example, assuming that 	$H_m$ is diagonalizable and not pathologically non-normal, we can cheaply compute its  eigenpairs $\theta_i\in \CC$, $\mathsf{s}_i\in\CC^N$ ($\mathsf{s}_i^T\mathsf{s}_j=\delta_{i,j}$) and  use spectral Lanczos decomposition
\begin{equation}\label{SDT} \mathsf{f} (t,H_m)e_1=\sum_{i=1}^m \mathsf{s}_i \mathsf{f} (t,\theta_i)(\mathsf{s}_i^Te_1).\end{equation}

Finally, we should point out that there is some cost associated with multiplication of matrix $V_m$ by vector $\mathsf{f}(t,H_m)e_1$  in the execution of (\ref{subapproxL}) and $V_m$'s storage, especially for large enough $m$.  However, this cost can be significantly reduced if the solution is only needed at a small subset of grid nodes. In this case (quite common for many applications) one needs just to use $V_m$'s submatrix consisting of the rows that correspond to such a subset.

\section{Electromagnetic wave propagation in a two-dimensional configuration}
\label{sec:em_2D} 

\begin{figure}
\begin{center}
\includegraphics*[width=0.45\textwidth]{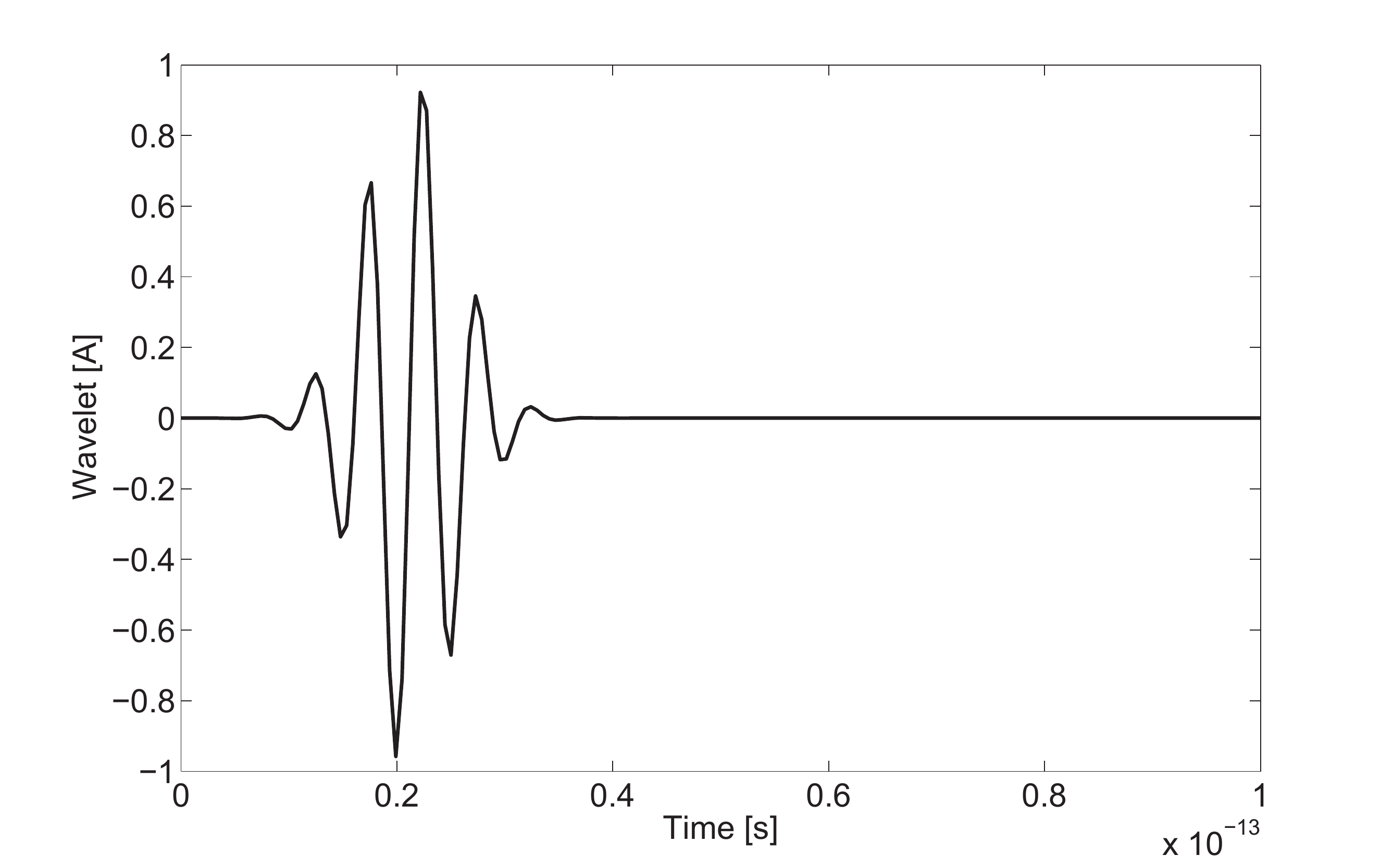}\includegraphics*[width=0.45\textwidth]{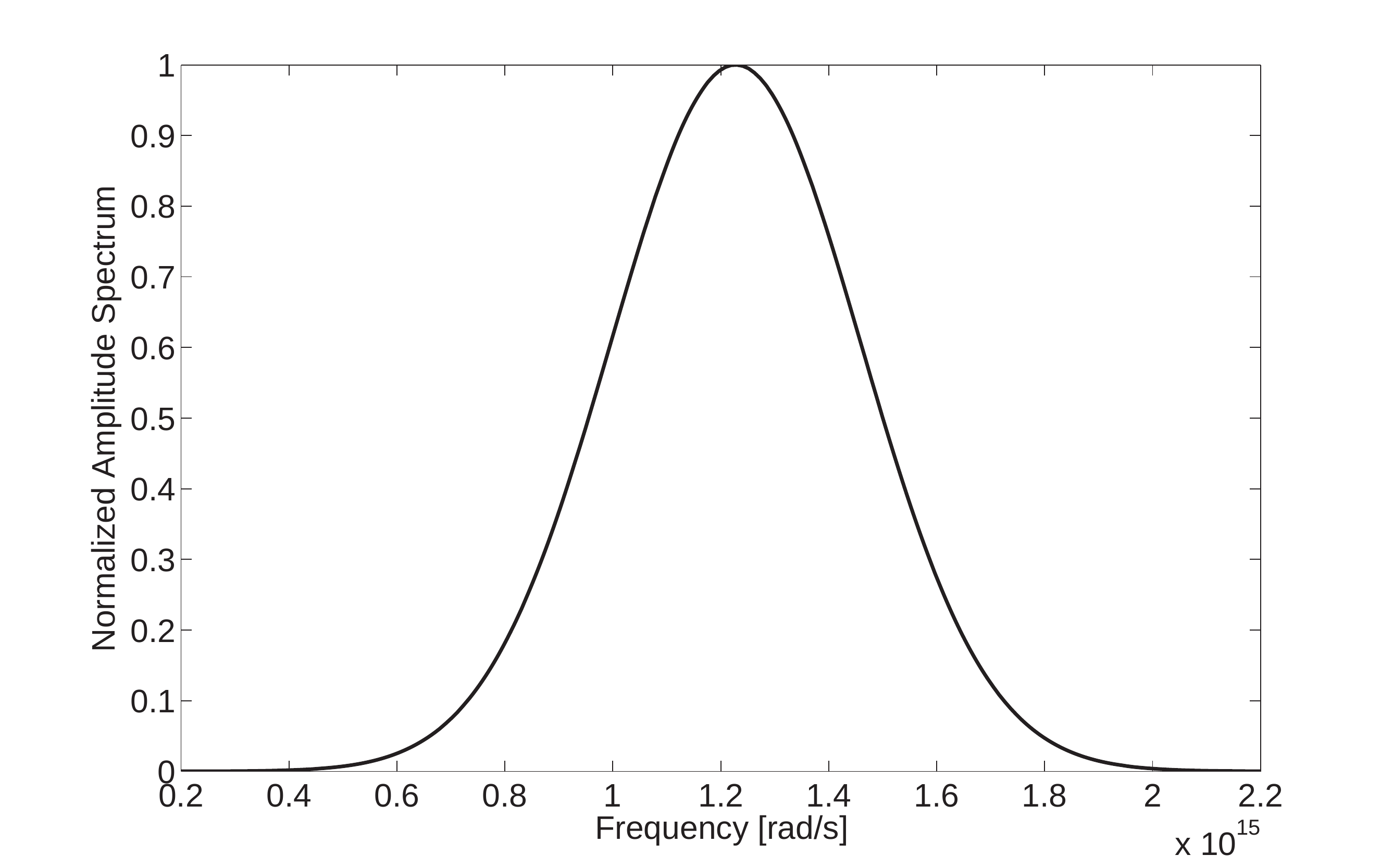}
\end{center}
\caption{Time signature $q(t)$ of the external electric-current source (left) and  normalized amplitude of its Fourier transform (right).}
\label{fig:wavelet&spectrum}
\end{figure}

To illustrate the performance of the stability-corrected spectral Lanczos method, we present some numerical experiments for E-polarized electromagnetic wavefields in two-dimensional configurations. By normalizing Maxwell's equations (with respect to a problem related reference length and the electromagnetic wave speed in vacuum) and by eliminating the magnetic field strength from the resulting set of equations, we end up with the wave equation (\ref{eq:swec}) for the electric field strength with $c=\varepsilon_{\text{r}}(x)$, where $\varepsilon_{\text{r}}(x)$ is the variable (time-independent) relative permittivity. 

In all experiments, electromagnetic waves are generated by an external electric-current source with a modulated Gaussian pulse~$q(t)$ as its time signature. 
In the first two sets of experiments, this modulated Gaussian has its spectrum in a frequency band~$[\omega_{\text{min}},\omega_{\text{max}}]$ with $\omega_{\text{min}}=2.42 \cdot 10^{14}$~rad/s and $\omega_{\text{max}}=2.18 \cdot 10^{15}$~rad/s (see Figure~\ref{fig:wavelet&spectrum}), while in the last set of experiments the pulse is tuned to a photonic waveguide structure for which $\omega_{\text{min}}=9.81 \cdot 10^{14}$~rad/s and $\omega_{\text{max}}=1.44 \cdot 10^{15}$~rad/s.   

To validate the results obtained with stability-corrected Lanczos, we compare computed field responses with analytic solutions or field responses obtained via a standard Auxiliary-Differential Equation PML implementation of the Finite-Difference Time-Domain method (ADE-FDTD method) with cubic polynomial PML profiles included~(for details, see \cite{Taflove_etal}). To make a fair comparison between both methods, we have implemented stability-corrected Lanczos in first-order form, since FDTD is based on the first-order Maxwell system as well. Furthermore, in each FDTD experiment the time step is set equal to the Courant upper limit. In the domain of interest $\Omega$, spatial discretization is identical in both FDTD and Lanczos codes. Specifically, discretization is chosen such that we have about 18 points per $\lambda_{\text{min}}$, where $\lambda_{\text{min}}$ denotes the smallest wavelength in the domain of interest that corresponds to the maximum frequency $\omega_{\text{max}}$. This leads to fully discretized interior domains with a few hundred step sizes in each Cartesian direction. For the discretized photonic waveguide problem, for example, we have 470 step sizes in each Cartesian direction. Finally, in all experiments we have used a five layer FFPML in the stability-corrected Lanczos approach and a ten layer PML was adopted in the FDTD method.  
\begin{figure}
\begin{center}
\includegraphics*[width=0.95\textwidth]{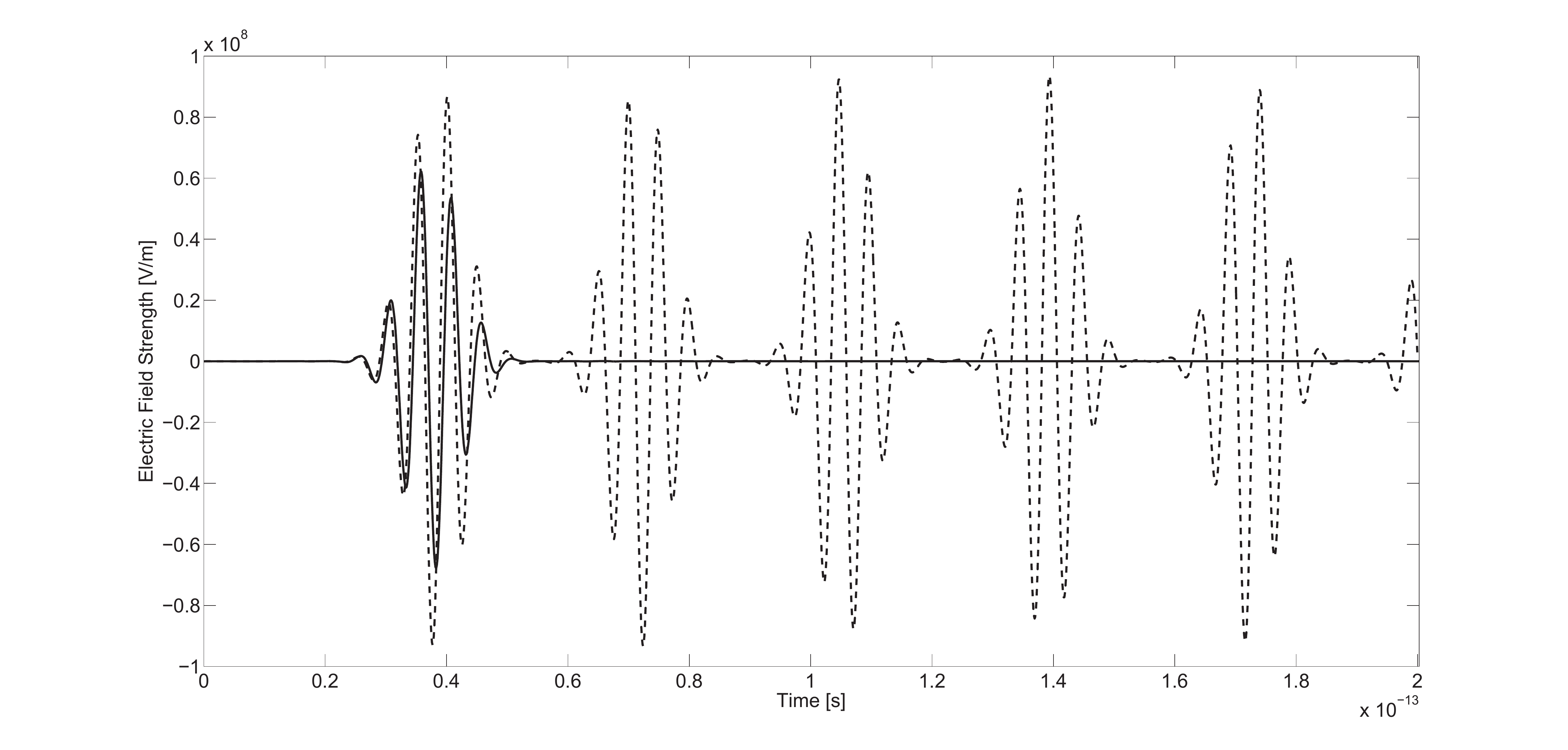}
\end{center}
\begin{center}
\includegraphics*[width=0.95\textwidth]{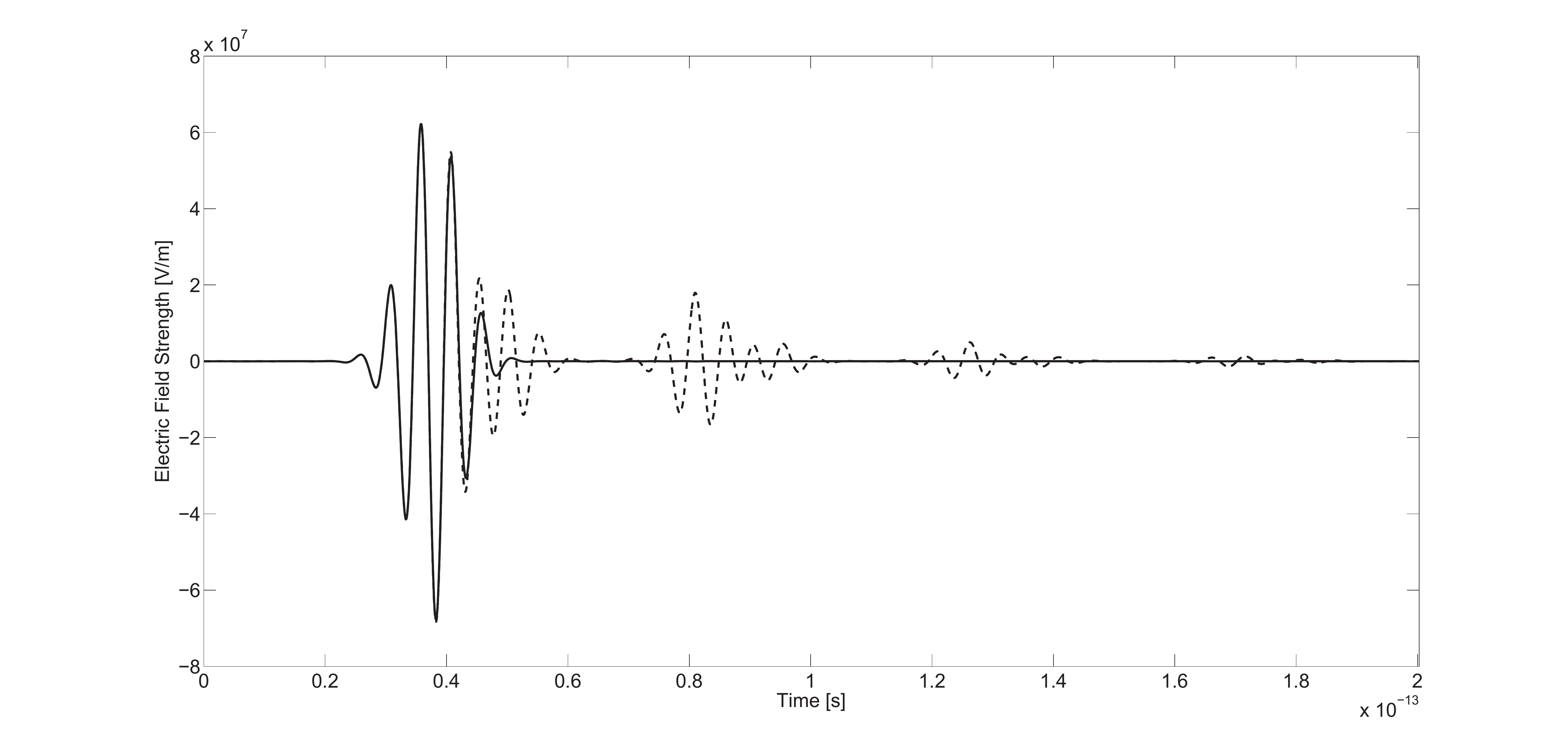}
\end{center}
\begin{center}
\includegraphics*[width=0.95\textwidth]{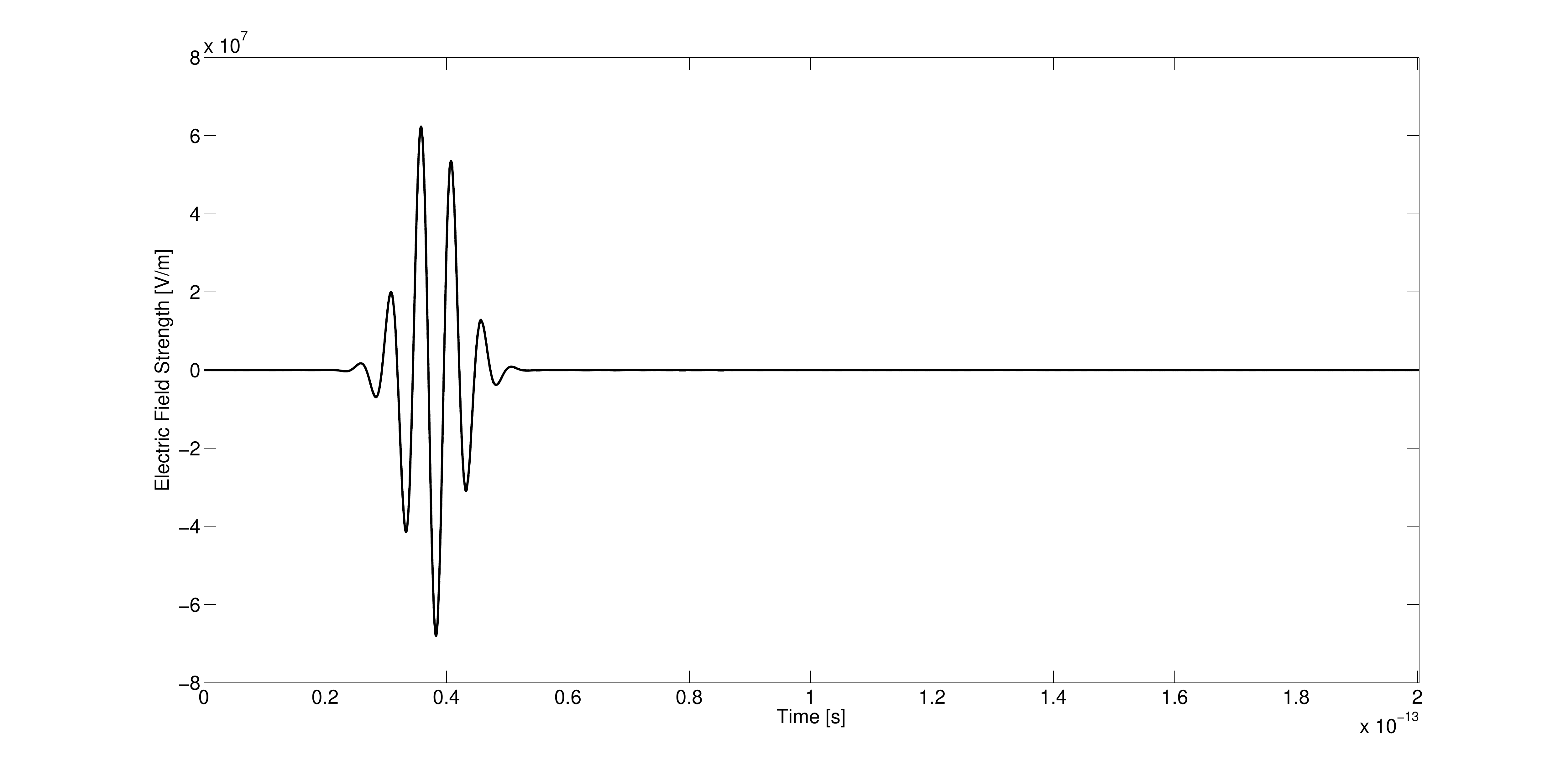}
\end{center}
\caption{Electric field strength at the receiver location on the time interval of interest. Solid line: analytic solution, dashed line: stability-corrected Lanczos after 300~(top), 400~(middle), and 500~(bottom) iterations.}
\label{fig:results_homogeneous}
\end{figure}
\subsection{Homogeneous Domain}
\label{subsec:HomogeneousDomain}
In our first set of experiments, we place the source in a vacuum domain and position the receiver at a distance $3\lambda_{\text{mid}}$ away from the source, where $\lambda_{\text{mid}}$ is the wavelength corresponding to the midfrequency $(\omega_{\text{min}}+\omega_{\text{max}})/2$. The electric field strength at the receiver location is computed via stability-corrected Lanczos method and we compare our results with the analytic solution for this problem. The time interval of interest runs from $t=0$~s to $t=2 \cdot 10^{-13}$~s. The solid line in Figure~\ref{fig:results_homogeneous} shows the analytic result on this time interval. Also shown are the computed field responses obtained with the spectral Lanczos method after 300 (Figure~\ref{fig:results_homogeneous}, top), 400 (Figure~\ref{fig:results_homogeneous}, middle), and 500 (Figure~\ref{fig:results_homogeneous}, bottom) iterations. The latter result coincides with the analytic result on the complete time interval of interest. 

\subsection{Dielectric Ring}
\label{subsec:DielectricRing}
\begin{figure}
\begin{center}
\includegraphics*[width=0.6\textwidth]{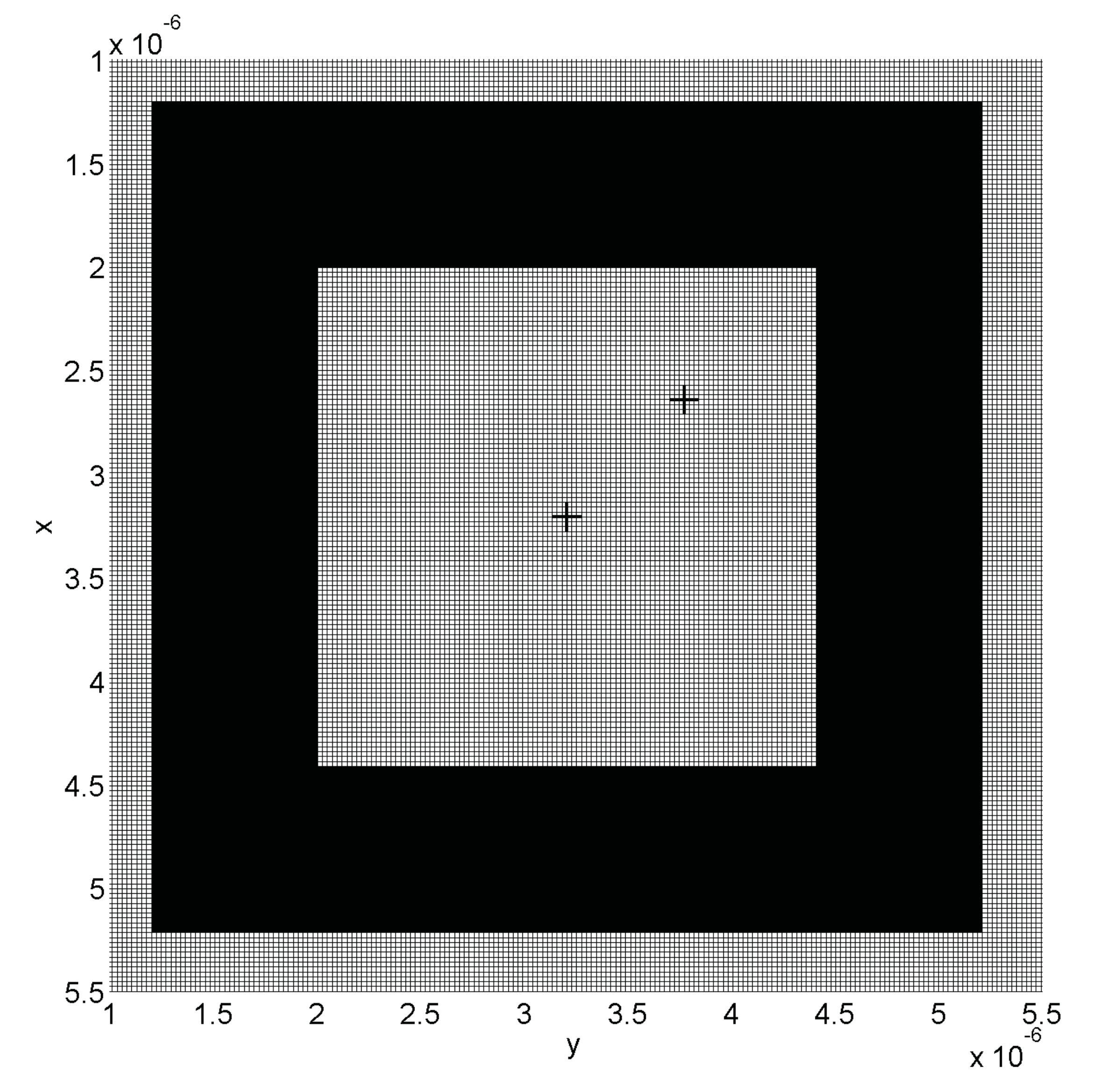}
\end{center}
\caption{A dielectric ring with a relative permittivity~$\varepsilon_{\text{r}}=4$ embedded in vacuum. The source is located at the center of the ring (plus sign in the middle of the figure). The plus sign located to the north-east of the source indicates the location of the receiver.}
\label{fig:ring}
\end{figure}

\begin{figure}
\begin{center}
\includegraphics*[width=0.95\textwidth]{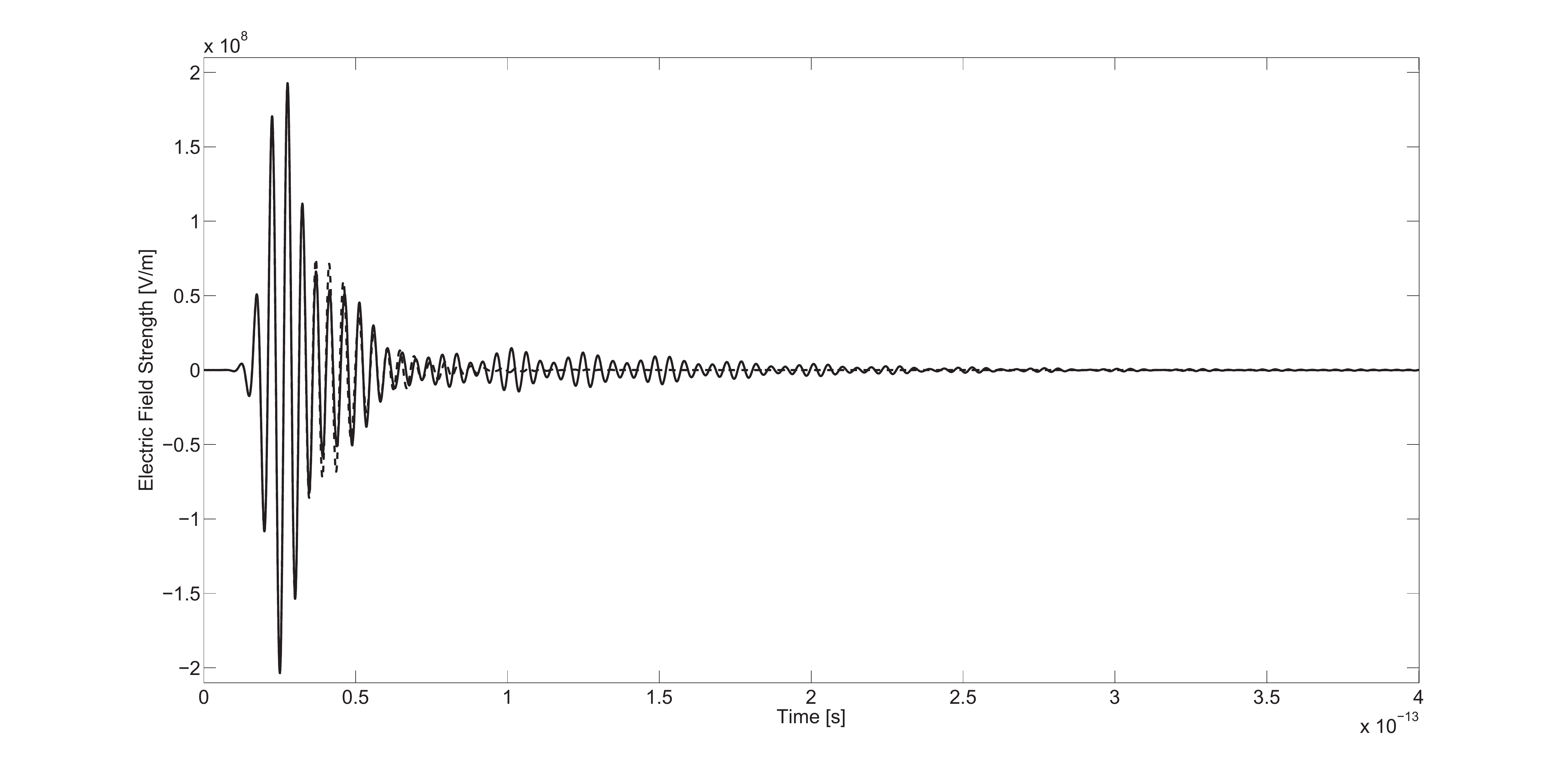}
\end{center}
\begin{center}
\includegraphics*[width=0.95\textwidth]{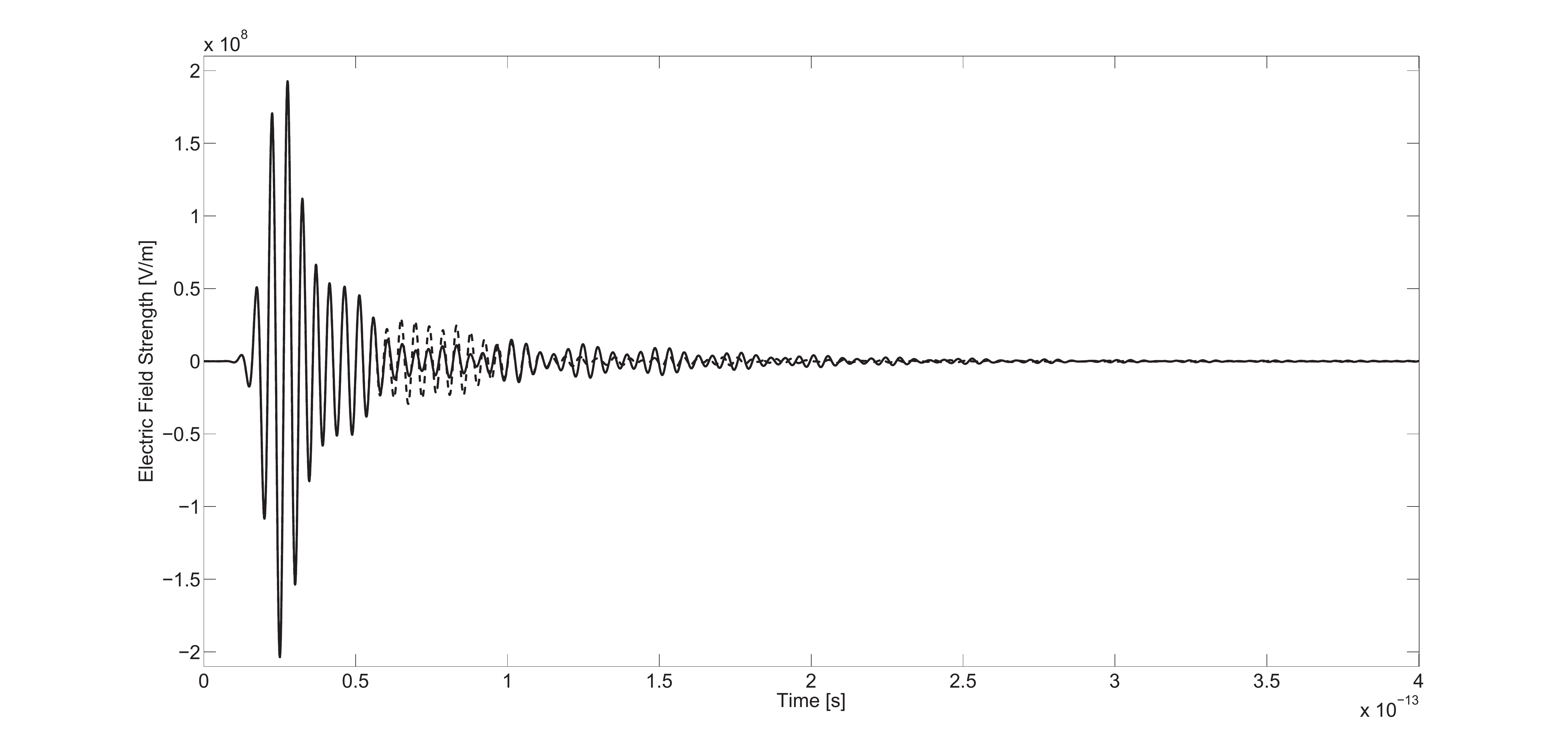}
\end{center}
\begin{center}
\includegraphics*[width=0.95\textwidth]{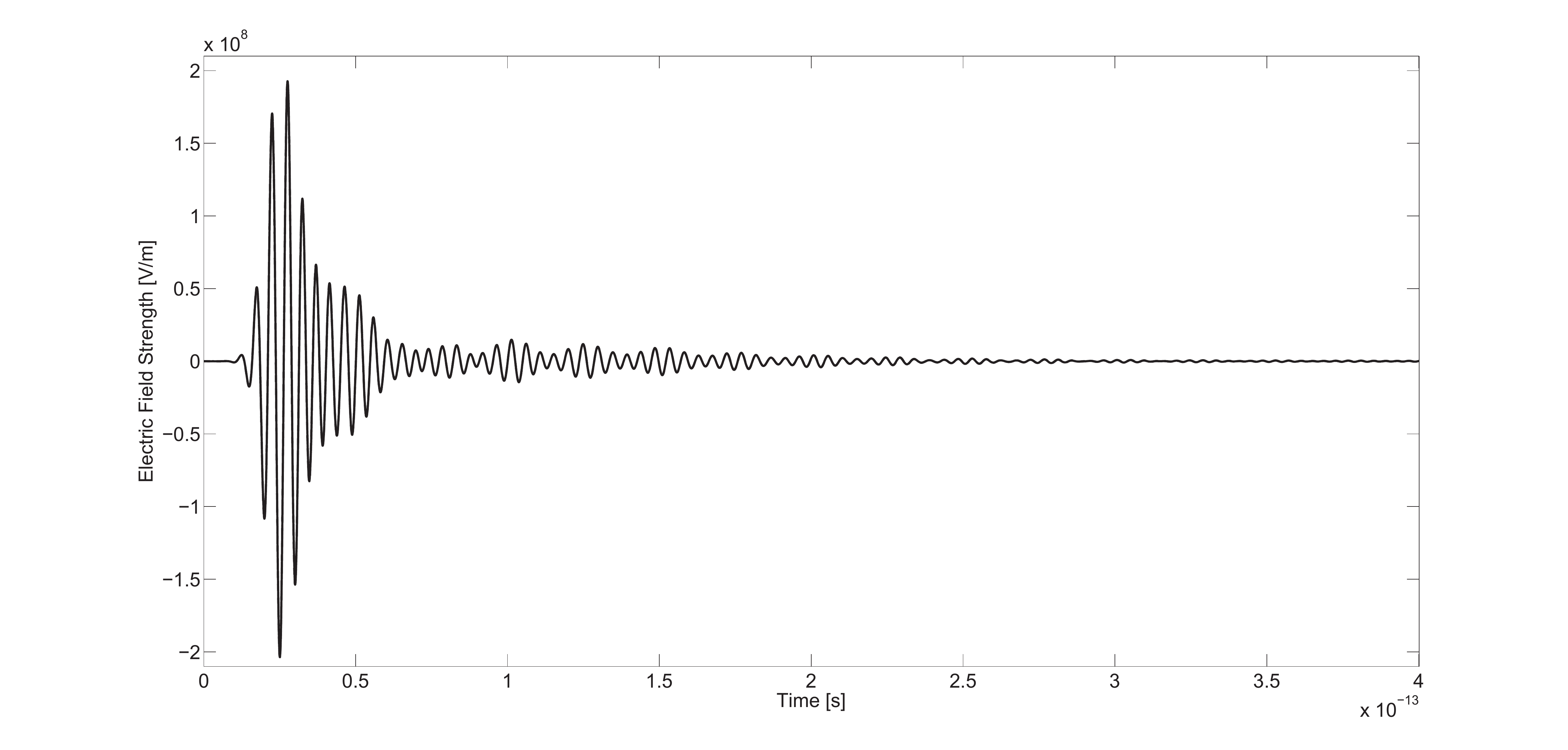}
\end{center}
\caption{Electric field strength at the receiver location on the time interval of interest. Solid line: FDTD after 7194~iterations, dashed line: stability-corrected Lanzos after 1000~(top), 2000~(middle), and 4000~(bottom) iterations.}
\label{fig:results_ring}
\end{figure}

\begin{figure}
\begin{center}
\includegraphics*[width=0.95\textwidth]{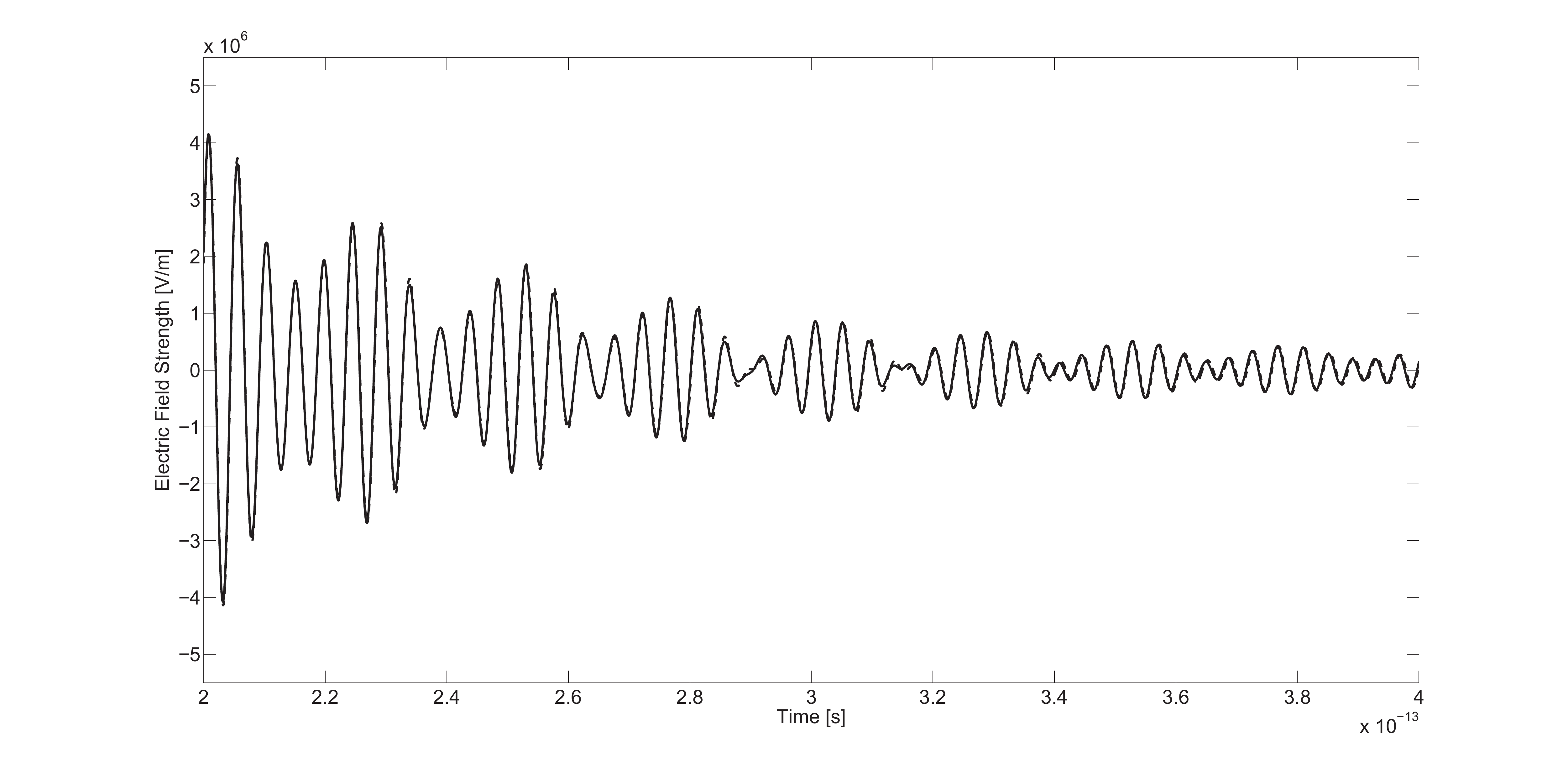}
\end{center}
\caption{Electric field strength at the receiver location on the second half of the time interval of interest. Solid line: FDTD after 7194~iterations, dashed line: stability-corrected Lanczos after 4000~iterations.}
\label{fig:results_ring_zoom}
\end{figure}

In our second set of experiments, we consider a dielectric ring with a relative permittivity $\varepsilon_{\text{r}}=4$ embedded in vacuum (see Figure~\ref{fig:ring}). Both the source (plus sign at the center of Figure~\ref{fig:ring}) and the receiver are located in the middle empty part of the ring. The time interval of observation runs from $t=0$~s to $t=4 \cdot 10^{-13}$~s and we compute the electric field strength at the receiver location by FDTD and the stability-corrected Lanczos method. The solid line in Figure~\ref{fig:results_ring} shows the response as computed by the FDTD method. For this problem, FDTD requires 7194~iterations to reach the end of the observation interval. The dashed line in the Figure~\ref{fig:results_ring}~(top) shows the Lanczos response on the time interval of interest obtained after 1000~iterations. Clearly, there is very little overlap with FDTD. The computed field responses improve, however, if we increase the number of Lanczos iterations. After 2000~iterations we obtain the result as shown in Figure~\ref{fig:results_ring}~(middle) and after 4000~iterations the computed electric field strength almost completely overlaps with the computed FDTD response, see Figure~\ref{fig:results_ring}~(bottom). In Figure~\ref{fig:results_ring_zoom} we zoom in on the second half of the observation interval to show that the Lanczos approximation of order~4000 has indeed almost converged to the FDTD field response. In Table~\ref{tab:comp_times}, we summarize the computation times that were required to finish the 4000 Lanczos iterations and 7194 FDTD iterations. Both methods were implemented in Matlab and computation times were measured on a computer with an Intel~Core~i7 Q740 CPU running at 1.73~GHz.  
\begin{figure}
\begin{center}
\includegraphics*[width=0.6\textwidth]{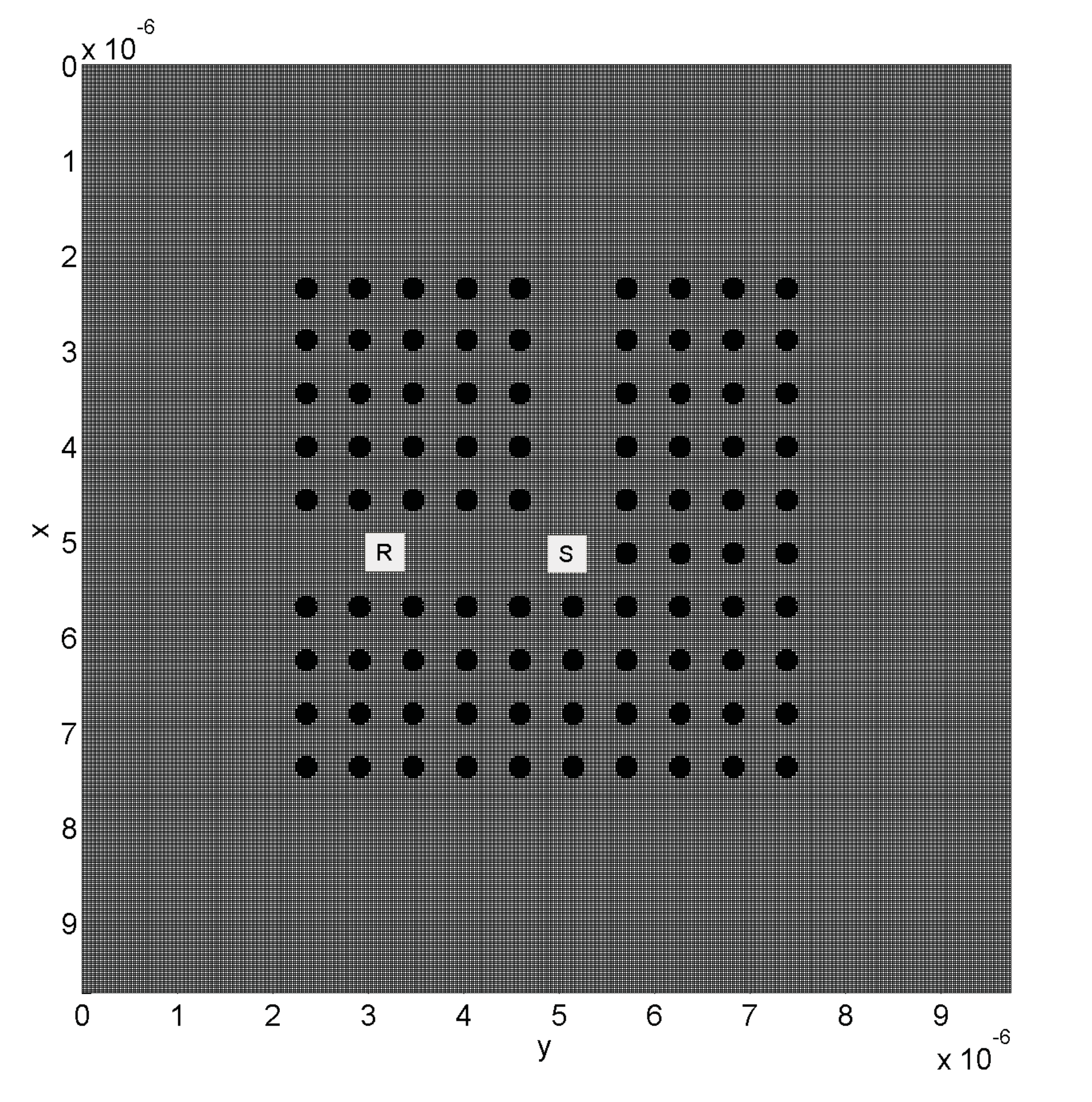}
\end{center}
\caption{A photonic waveguide structure consisting of dielectric rods with a relative permittivity of $11.56$. The distance between the rods is $\ell=0.58~\mu$m and the radius of the rods is 
$0.18\ell$. The letters S and R indicate the location of the source and the receiver, respectively.}
\label{fig:crystal}
\end{figure}
\begin{figure}
\begin{center}
\includegraphics*[width=0.95\textwidth]{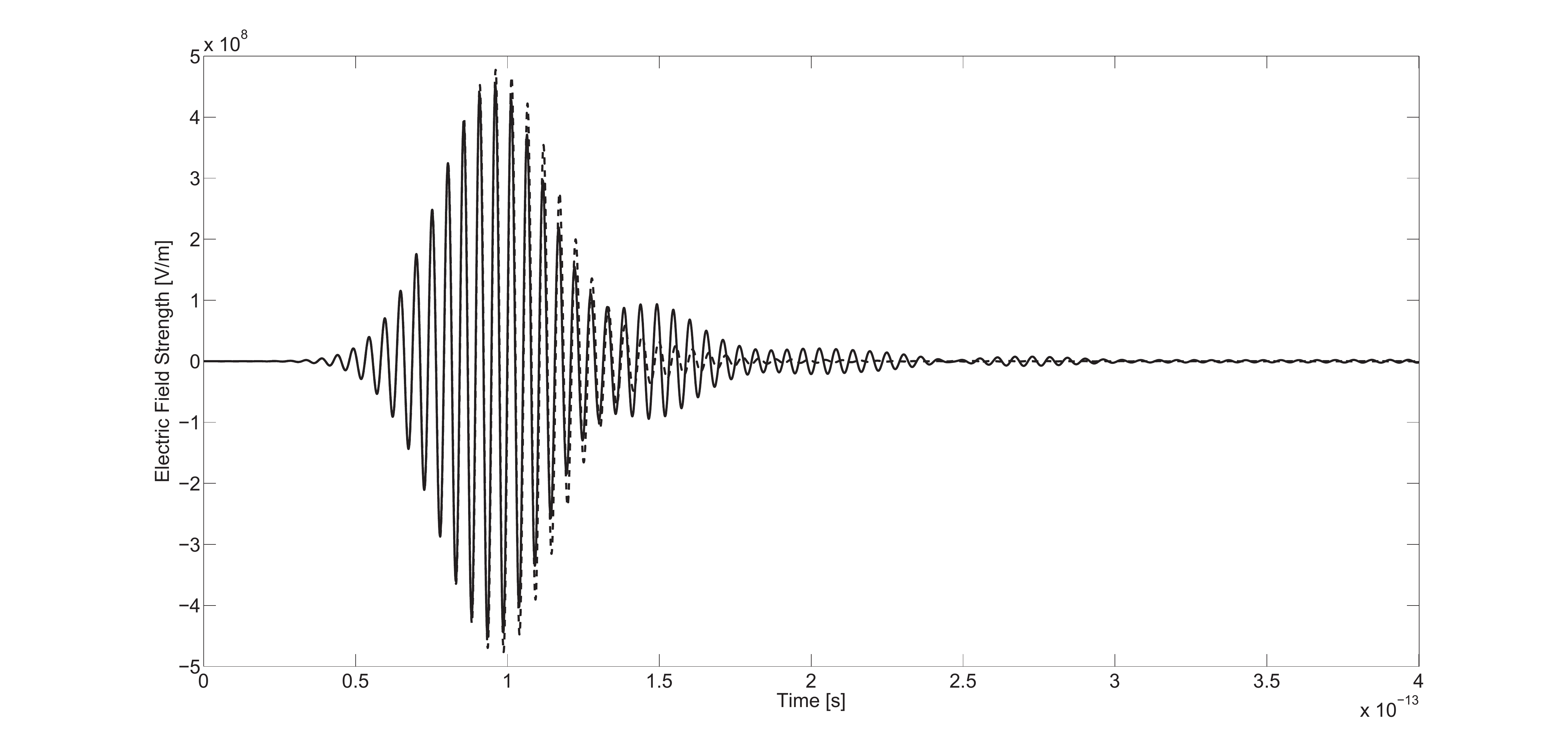}
\end{center}
\begin{center}
\includegraphics*[width=0.95\textwidth]{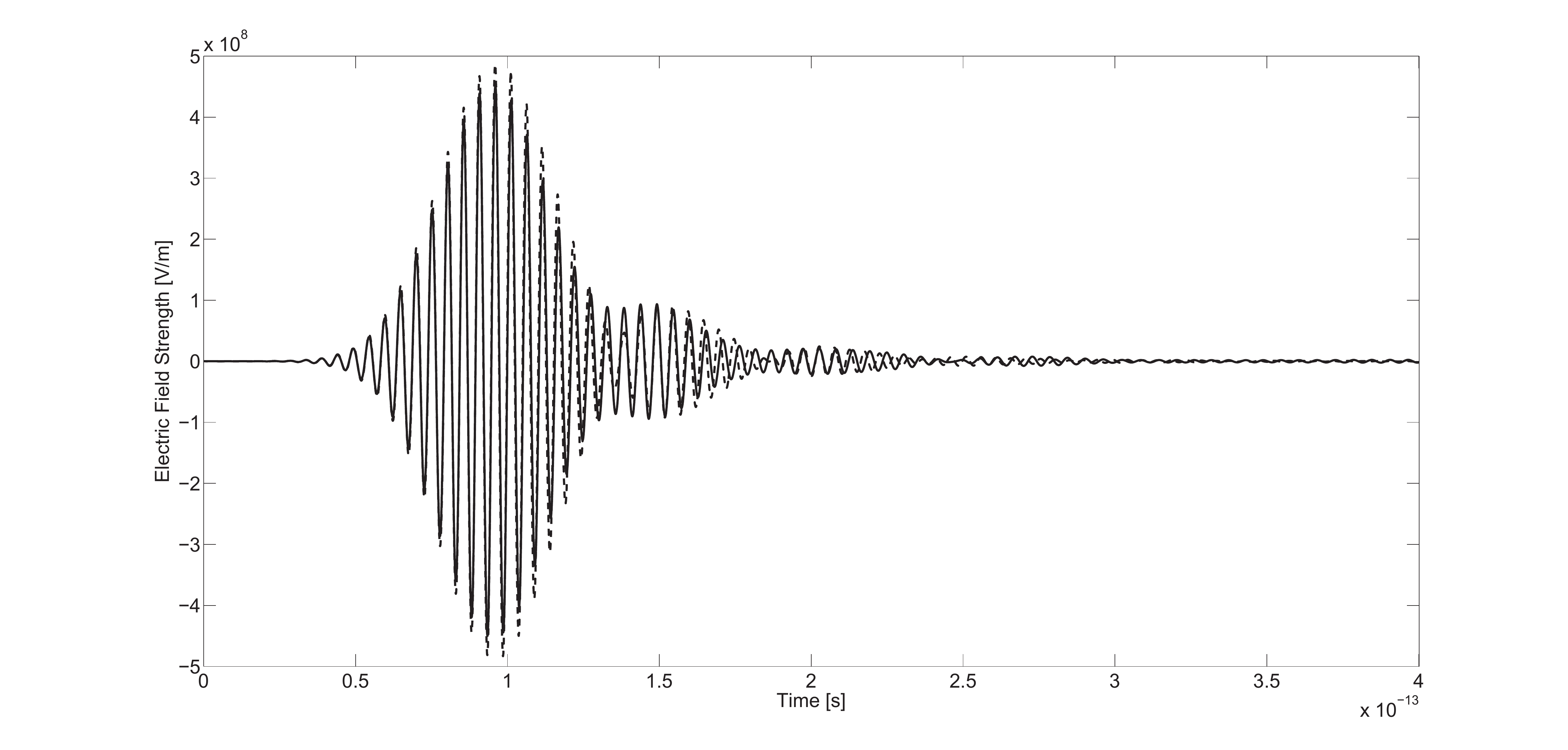}
\end{center}
\begin{center}
\includegraphics*[width=0.95\textwidth]{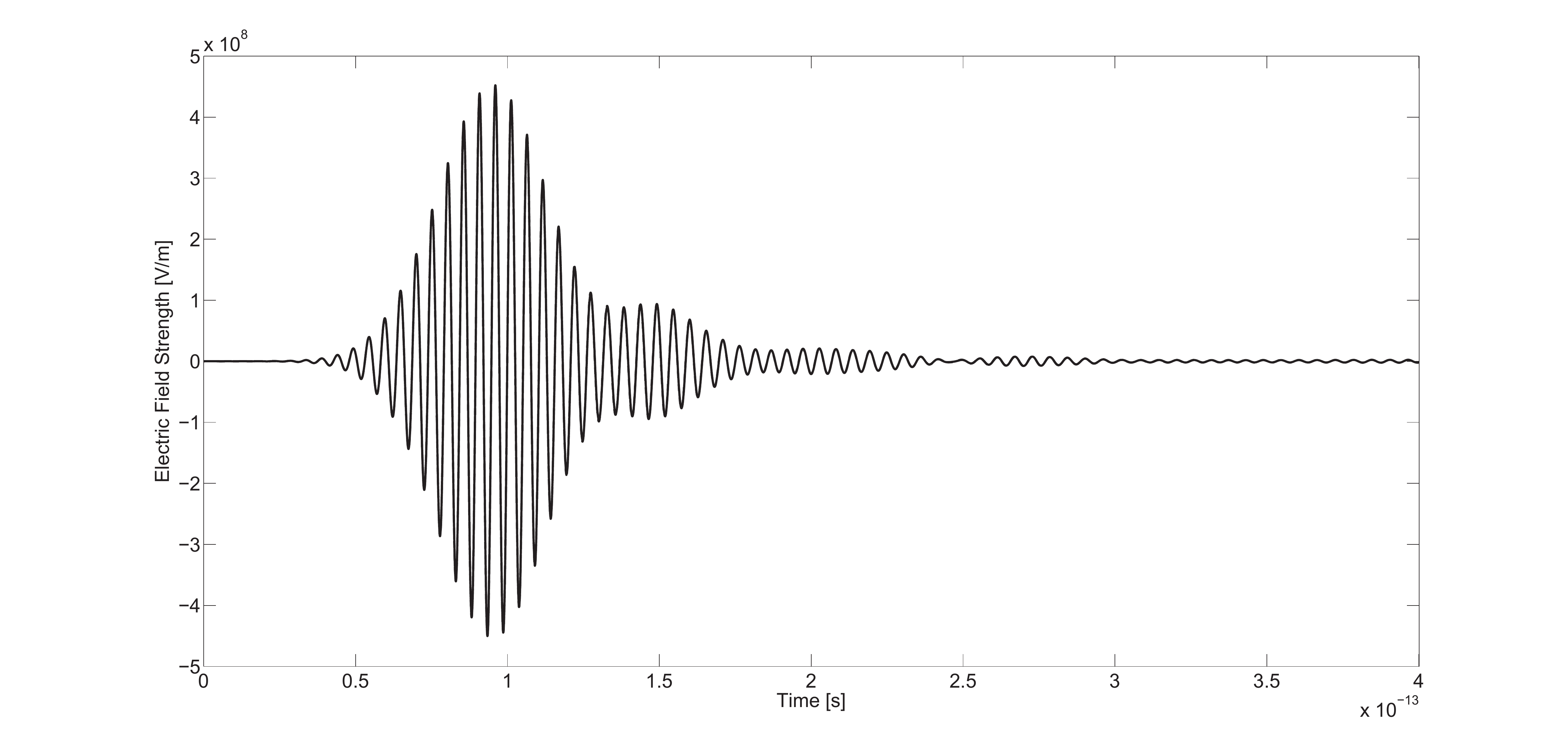}
\end{center}
\caption{Electric field strength at the receiver location on the time interval of interest. Solid line: FDTD after 8197 iterations, dashed line: stability-corrected Lanczos after 1000 (top), 2000 (middle), and 3000 (bottom) iterations.}
\label{fig:results_crystal}
\end{figure} 
\begin{figure}
\begin{center}
\includegraphics*[width=0.95\textwidth]{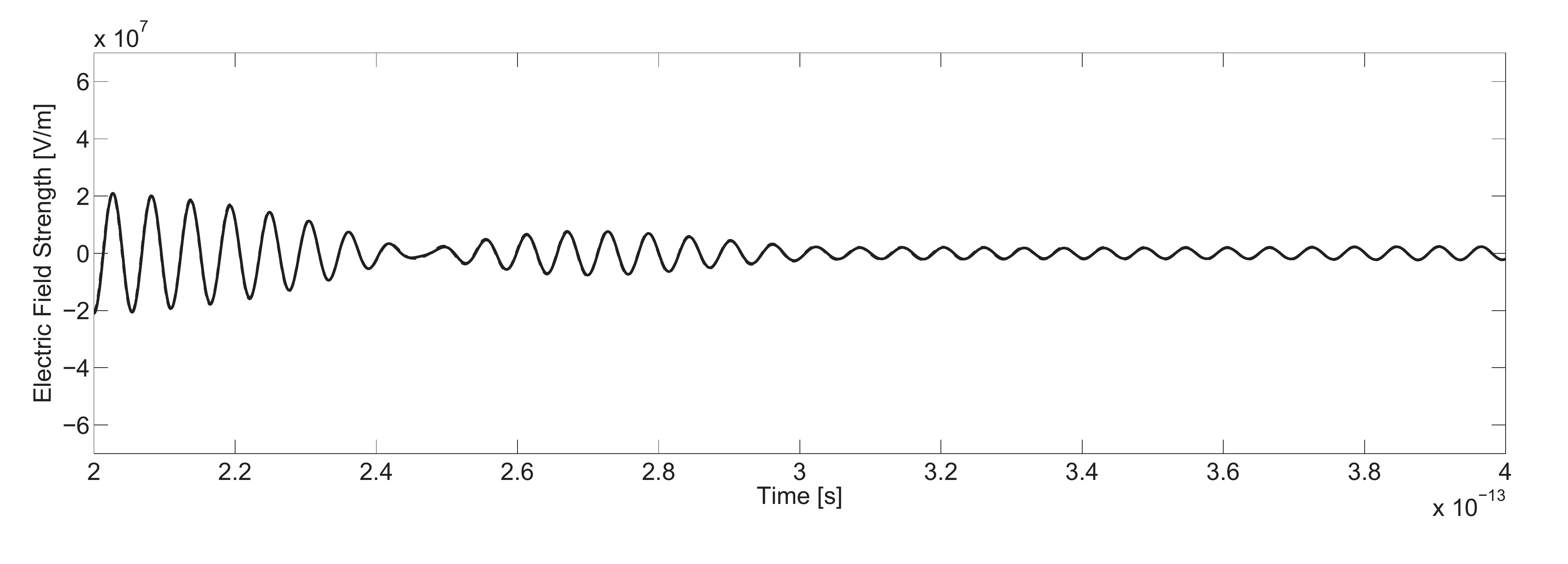}
\end{center}
\caption{Electric field strength at the receiver location on the second half of the time interval of interest. Solid line: FDTD after 8197~iterations, dashed line: stability-corrected Lanczos after 3000~iterations.}
\label{fig:results_crystal_zoom}
\end{figure}
\subsection{Photonic Waveguide}
\label{subsec:crystal}
In our final set of experiments, we consider the simple photonic waveguide structure shown in Figure~\ref{fig:crystal}. This structure is modeled after a photonic crystal presented in \cite{Mekis_etal} and consists of a set of dielectric rods placed in vacuum. The distance~$\ell$ between the rods is approximately 0.58~$\mu$m and each rod has a radius of $0.18\ell$. The relative permittivity of the rods is 11.56.  

By removing half a row and half a column of rods, a bend is introduced inside the crystal. The source is positioned at the corner of the bend (see Figure~\ref{fig:crystal}) and since the wavelet has its spectrum in the bandgap of the crystal (see \cite{Mekis_etal}), electromagnetic waves will propagate to the left and to the top of the crystal along the artificially created photonic waveguide structure.   
We compute the electric field strength at a position approximately halfway one of the waveguides (see Figure~\ref{fig:crystal}). In Figure~\ref{fig:results_crystal} we show the responses computed by FDTD and stability-corrected Lanczos. The solid line again shows the FDTD response on a time interval of observation that runs from $t=0$~s to $t=4 \cdot 10^{-13}$~s. It takes FDTD 8197~iterations at the Courant limit to reach the end of the observation interval. The dashed line in Figure~\ref{fig:results_crystal}~(top) shows the result obtained after 1000 Lanczos iterations. There is no good agreement with FDTD yet. Increasing the number of iterations to 2000, we obtain the result as shown in Figure~\ref{fig:results_crystal}~(middle) and after 3000~iterations we obtain a response as signified by the dashed line in Figure~\ref{fig:results_crystal}~(bottom). The latter result overlaps with the FDTD result on the complete time interval of observation as can also be seen in Figure~\ref{fig:results_crystal_zoom}, where we show the field response on the second half of the observation interval. The computation times that were required to finish the 3000 Lanczos iterations and 8197 FDTD iterations are summarized in Table~\ref{tab:comp_times}. 

\subsection{Conclusions}
\label{subsec:concl}
   
The numerical experiments show that the stability corrected Lanczos algorithm converges rather uniformly on the entire spectral interval. Convergence is achieved if the part of $\TAN$'s spectrum that is closest to the real axis is well approximated. In this case, the algorithm gives an accurate leading term of the late time scattering pole asymptotes described in \cite{LaxPhillips, TangZworski}. By contrast, the FDTD cost is just strictly proportional to the propagation time interval. On the one hand, this implies that FDTD would be more economical if we made comparisons on much smaller time intervals. On the other hand, we could have achieved much larger speedups than reported here, had we carried out our benchmarks on larger time intervals. We conclude therefore that the stability corrected Lanczos algorithm is much more efficient than the FDTD for exterior wave problems with large time intervals of observation, see below  for further discussion.  

\begin{table}
\centering
\caption{Number of iterations (NOI) and corresponding computation times (CT) in minutes for the dielectric ring and photonic waveguide problem.} 
\begin{tabular}{|l|c|c|c|c|}
\hline
\hline
Problem & \multicolumn{2}{|c|}{Lanczos} & \multicolumn{2}{c|}{FDTD} \\ \hline
 & NOI & CT (min) & NOI  & CT (min) \\ \hline
Ring & 4000 & 3.8 & 7194 & 5.7 \\ \hline
Waveguide & 3000 & 10.1 & 8197 & 29.4 \\ \hline
\end{tabular}
\label{tab:comp_times}
\end{table}

\section{Concluding remarks}
\label{sec:concl_rem}
 
\begin{itemize}

\item Numerical experiments (see conclusion of the previous section) clearly show that the (polynomial) Krylov subspace SCTDE algorithm bypasses the convergence threshold of matrix polynomial methods for wave problems in selfadjoint formulations as obtained in \cite{DK89}. However, this advantage dissappears and even reverses for small time intervals. This can be explained by the appearance of a square-root singularity in the non-selfadjoint SCTDE formulation. It is known that rational Krylov subspaces (RKSs)\cite{Druskin&Knizhnerman98,BR2008} can efficiently handle matrix functions with such singularities. The framework developed here allows for a generalization to these RKS methods.
\item Unconditional stability of the SCTDE is a property that makes our approach especially attractive for problems with known unstable PMLs, e.g., layered elasticity problems including fully anisotropic media \cite{Becache_etal,SavadattiGuddati}.
\item The stability-corrected resolvent given by (\ref{SKR}) has  spectral properties similar to the ones of  the true resolvent (\ref{eq:solLaplace}) at least
on the main Riemann sheet, i.e., the both are analytic functions of $\lambda$ on $\CC\setminus (-\infty,0)$ with the branch cuts on $(-\infty,0)$. 
$\TAN$'s spectrum was successfully used in \cite{Bindel} for identification of the true scattering poles in some neighborhood of $-{\omega_0}^2$. Thus, we expect, that in our case at least
well separated  singularities of the stability corrected resolvent  (\ref{SKR}) approximate well separated  scattering poles of the exact problem, and the rest gives some integral approximation.  In that light  we possibly can view spectral decompositions (\ref{SD}) and (\ref{SDT}) as  approximative  counterparts of the asymptotic resonance expansions of  \cite{LaxPhillips,TangZworski}. Such a connection is worth a special investigation, of course.
\item When we were preparing this manuscript, Leonid Knizhnerman showed us that the SCTDE presented here is not the only matrix function of damped operators allowing stable time-domain computations, and possibly there exists a class of such solutions.  

\end{itemize}
\section{Acknowledgments}
We  thank our friends and colleagues Leonid Knizhnerman, Olga Podgornova, and Mikhail Zaslavsky for their careful reading of a preliminary draft of this paper, making some useful suggestions, 
corrections, and preliminary calculations that verified some of the results of this work. We are also indebted to Robert Kohn for bringing work \cite{TangZworski} to our attention.

\end{document}